\documentclass[12pt,preprint]{aastex}
\usepackage{here}
\usepackage{amsmath}
\usepackage{amsfonts}
\usepackage{amssymb}
\usepackage{graphicx}
\usepackage[squaren,Gray]{SIunits}
\usepackage{xcolor}
\usepackage{fancyhdr}
\usepackage{longtable}
\usepackage{lscape}
\usepackage{multirow}
\usepackage{url}
\usepackage{natbib}
\usepackage{vmargin}

\begin{document}

\title{Mass distributions of stars and cores in young groups and clusters}
\author{Manon Michel}
\affil{Formation Interuniversitaire de Physique, D\'epartement de Physique de l'\'Ecole Normale
Sup\'erieure, 24 rue Lhomond, 75231 Paris Cedex 05, France;\email{manon.michel@ens.fr}}
\author{Helen Kirk}
\affil{Harvard-Smithsonian Center for Astrophysics, 60 Garden Street, MS 42, Cambridge MA 02138, USA \email{hkirk@cfa.harvard.edu}}
\and
\author{Philip C. Myers}
\affil{Harvard-Smithsonian Center for Astrophysics, 60 Garden Street, MS 42, Cambridge MA 02138, USA \email{pmyers@cfa.harvard.edu}}

This is an author-created, un-copyedited version of an article accepted for publication in Astrophysical Journal. IOP Publishing Ltd is not responsible for any errors or omissions in this version of the manuscript or any version derived from it.

\begin{abstract}
We investigate the relation of the stellar initial mass function (IMF) and the dense core mass function (CMF), using stellar masses and positions in 14 well-studied young  groups. Initial column density maps are computed by replacing each star with a model initial core having the same star formation efficiency (SFE).  For each group the SFE,  core model, and observational resolution are varied to produce a realistic range of initial maps.  A clumpfinding algorithm parses each initial map into derived cores, derived core masses, and a derived CMF.  The main result is that projected blending of initial cores causes derived cores to be too few and too massive. The number of derived cores is fewer than the number of initial cores by a mean factor 1.4 in sparse groups and 5 in crowded groups.  The mass at the peak of the derived CMF exceeds the mass at the peak of the initial CMF by a mean factor 1.0 in sparse groups and 12.1 in crowded groups. These results imply that in crowded young groups and clusters, the mass distribution of observed cores may not reliably predict the mass distribution of protostars which will form in those cores.

\end{abstract}

\section*{Introduction}
\par For a star, its mass is the most important parameter regarding its evolution and interaction with environment. On larger scales, galaxy formation and evolution are also shaped by the mass distribution of stars. Moreover the initial star mass distribution, the Initial Mass Function (IMF), appears strikingly to be universal with a power law behaviour for masses above 1 $M_\odot$ \citep{Kroupa2002} and a lognormal shape below \citep{Chabrier2000}. Therefore, the IMF is a key issue and a compulsory theory test in star formation, one of the most basic but still most challenging problems in astrophysics.
\par It is now well established that star formation occurs in the densest and coldest parts of molecular clouds, known as dense cores (see \cite{Ward-Thompson1994}, \cite{Kirk2005}, \cite{Ward-Thompson2007}). As cores set the initial conditions for star formation, the core mass function (CMF) is often compared to IMF to provide constraints on star formation efficiency, timescale and fragmentation. In addition to core density and temperature structures, the CMF may also be a theory test by comparing observed and expected CMFs. But above all, the CMF could answer the still unsolved question of fragmentation, especially regarding massive star formation. The presumed differences between CMFs derived for example from competitive accretion model \citep{Bonnell1997}, monolithic collapse \citep{McKee2003} or some in-between models could indeed distinguish these theories. If a strong link between CMF and IMF is found, the IMF origin issue may be answered by directly solving the CMF one.
\par Recent CMF measurements in low-mass star forming regions have shown a great likeness in shape between IMF and CMF (\cite{Motte1998} and \cite{Johnstone2000} in $\rho$ Ophiuchi main cloud, \cite{Alves2007} and \cite{Rathborne2009} in Pipe Nebula, \cite{Testi1998} in Serpens core, \cite{Motte2001} and \cite{Johnstone2000} in Orion B, \cite{Tothill2002} in the Lagoon nebula). These observations confirm the role played by cores as direct precursors for stars. But it also strengthens the idea of a one-to-one relationship between individual stars and individual pre-stellar cores. As a consequence, it enhanced also the idea that the shift between the break or peak masses observed in IMF and CMF is due to a star formation efficiency less than 1. The star formation efficiency is the ratio between the final star mass and its direct progenitor core initial mass; the currently favoured value is about 1/3 \citep{Alves2007}.
\par The interpretation of the shift between the IMF and CMF as caused by the efficiency of star formation is being questioned, however. \cite{Swift2008} have shown through numerical experiments that the comparison between the mass functions of dense cores and stars might only distinguish  with great difficulties one among several evolutionary schemes, given the current observational accuracy. They add that the best way of distinguishing several schemes is to study the mass functions over a wide range of mass scales including the high- or low-mass tails. Moreover \cite{Goodwin2008} have argued that the CMF derived in \cite{Nutter2007} will best match the IMF derived in \cite{Kroupa2002} if cores are assumed to form binaries or higher-order multiples. Observational results in Perseus \citep{Hatchell2008} have provided some new insights into the possible evolution of a prestellar core. The mass distribution of prestellar and protostellar cores in this region appeared indeed inconsistent with a simple direct one-to-one mapping to the IMF.
\par In addition to the nature of the shift between IMF and CMF, the similarity in shape between the two distributions could also be interpreted as a consequence of the central limit theorem. It has been argued that the IMF is set through the act of a large number of independent physical processes and variables, which lead to its lognormal shape (\cite{Larson1973}, \cite{Zinnecker1984} and \cite{Adams1996}). Therefore, the CMF would also present a lognormal shape, as it is produced in a similar way. Moreover, the process of deriving a CMF from observational data adds independent random processes which differ from those used to obtain the IMF (these include random errors on core boundaries, core masses, cores blending which do not occur when objects are point sources as in IMF derivation) \citep[see][]{Reid2010}. The lognormal shape of both IMF and CMF would then be independent of a one-to-one relationship between prestellar cores and stars.
\par Those uncertainties about the exact relationship between the IMF and CMF appear to remain unresolved in a large part because the mass distribution of stars that will actually form eventually from an observed CMF is unavailable. In young stellar groups, however, where stars have not moved much from their birth sites, it is possible to approach numerically what natal cores looked like. The original dense core column density map can be estimated based on the current young stellar population, allowing the derived CMF to be directly compared to the actual local star distribution.
\par Starting from YSO distributions divided in groups by \cite{Kirk2010} in four nearby star-forming regions presenting different degrees of crowding, we recreated a possible original column density map for each group. Each YSO contributes to the global column density map of the group by adding one prestellar core profile centred at the YSO position. This profile encloses a total mass equal to the YSO mass, divided by a fixed star formation efficiency, either 0.3 or 1.0. We used two well-known profiles representative of different points of view on star formation, the Critically Stable Bonnor Ebert sphere (CSBE) (see \cite{Bonnor1956} and \cite{Ebert1955}) and the Thermal Non-Thermal sphere (TNT) (see \cite{Myers1992}). Mimicking the observational CMF derivation, we used the clumpfind algorithm clfind2d \citep{Williams1994} to obtain the core mass function and compare it directly to the local star mass distribution. From now on, we will address by `initial' CMF, the core mass distribution obtained from the local star mass distribution by a one-to-one relationship between cores and stars and by a fixed star formation efficiency; we will address by `derived' CMF, the mass distribution of cores identified by clfind2d on the simulated column density maps. The properties obtained from the `derived' CMF will  be also called `derived' (e.g. `derived' star formation efficiency).
\par Our main results are that the numbers of derived cores are undercounted and their masses overestimated, due to blending between cores, even in relatively  uncrowded regions. This happens with no added background and noise and no variation in the number of stars per initial core; the inclusion of these effects would increase the inconsistency between original and derived cores. We find also that the blending effects are so strong in crowded regions that they dominate other effects, such as differing core models used to derive the column density maps.
\par The data and method used are presented in more details in Section \ref{sec:Method}, then results are displayed in Section \ref{sec:Results} and discussed in Section \ref{sec:Discussion}. 
\section{Method}
\label{sec:Method}
\subsection{Initial YSO distributions}
\par This study is based on the 14 groups derived by \cite{Kirk2010} from catalogs of YSOs in 4 nearby regions. The regions were divided into groups by using the Minimum Spanning Tree (MST) algorithm \citep{Barrow1985}, following the procedure of \cite{Gutermuth2009}. The four analyzed nearby star-forming regions are Taurus, Chamaeleon I, Lupus 3 and IC 348 and all have catalogs extending to late M (or even L0) spectral types. The primary catalog used for each region is \cite{Luhman2010} (for Taurus), \cite{Comeron2008} (for Lupus 3), \cite{Luhman2007} (for Chamelaeon I) and \cite{Lada2006} plus \cite{Muench2007} (for IC 348). For the full catalogs, see \cite{Kirk2010}.
\par The distances adopted to each region are the same as in \cite{Kirk2010} : 140 pc for Taurus, 200 pc for Lupus 3, 160 pc for Cha I and 300 pc for IC 348. As stated in \cite{Kirk2010}, these four regions are the only ones within 300 pc containing stars younger than several million years and whose local extinction is low enough to make them observable in the optical/near-IR and have a spectral classification completeness of greater than 90 \%.
\par The mass estimation of YSOs was based on the spectral type procedure outlined in \cite{Luhman2003} using a combination of models from \cite{Palla1999}, \cite{Baraffe1998} and \cite{Chabrier2000}. The completeness limit is about 0.02 $M_\odot$ for all regions.
\par The MST algorithm used by \cite{Kirk2010} links all the YSOs of one region in a tree, ie a connected graph, which has the property of having the minimum total branch length. By using a cut-off length determined by the total distribution of branch lengths, the MST algorithm defines these groups in a way that mimicks the eye behaviour. See \cite{Kirk2010} for further details.
\par More details on the YSO spectral types, positions and mass estimation can be found in \cite{Kirk2010}.
\subsection{Going backward : From YSO to cores}
\subsubsection{Core models}
\par A column density map was derived from the locations and masses of the YSOs by using different core models and star formation efficiency values.
\par The two models used in this study are the Critically Stable Bonnor-Ebert Sphere model (CSBE) (see \cite{Bonnor1956} and \cite{Ebert1955}) and the Thermal-Non Thermal model (TNT model) (see \cite{Myers1992} and \cite{Myers2010}). We use these particular models as they are frequently used in the community and each offers a different point of view on star formation, in particular on the link between the mass enclosed and the clump radius. Their exact expressions can be found in Appendix \ref{appendix:models}.
\par The CSBE model was computed in two ways, which we label the CSBET model and the CSBEP model. For the CSBET model, the temperature was set to 16 K for all groups and the external pressure was allowed to vary. This temperature value is the same as the one measured in the Pipe nebula \citep{Alves2007}. For the CSBEP model, the external pressure was set to \textit{P/k} = 3.0x$10^{6}$ K.cm$^{-3}$ for all groups and the temperature was allowed to vary. This pressure value is thirty times the pressure measured in the Pipe nebula, as the Pipe nebula value was too low to describe most star forming regions. Typical star forming regions have higher values such as Ophiuchus (\cite[][]{Johnstone2000}) and Orion B \cite[][]{Johnstone2001}. The adopted pressure makes the core column density profile peaks at 30x$10^{21}$ cm$^{-2}$, which is consistent with high-resolution observations of nearby cores \citep[see][]{Alves2007,Kirk2006}. It is noteworthy that these pressure and temperature choices are consistent with the ones derived by assuming a Bonner-Ebert profile in \cite[][$\rho$ Ophiuchi]{Johnstone2000} (T $\approx$ 10-30K and \textit{P/k} $\approx 10^{6-7}$ K.cm$^{-3}$) and \cite[][Orion B]{Johnstone2001} (\textit{T} $\approx$ 20-40K and 3.$10^{5} \leq$ \textit{P/k} $\leq$ 3.$10^{6}$ K.cm$^{-3}$). For all models, we assumed a mean particle mass, m, of 2.3 times the mass of hydrogen.
\par Figure \ref{fig:model_profile} displays for each model the column density profiles versus the projected radius for an enclosed mass of 0.5, 1, 3, 5 and 10 $M_\odot$. Whereas the profiles of the CSBET and the TNT models are similar when the mass enclosed is low, they differ at high enclosed masses. From now on, we will describe the column density profiles in terms of `core' component (spiky component) and `clump' component (the extended halo surronding the centrally peaked `core' component). At low masses, the `core' component of the TNT model dominates and is essentially a singular isothermal sphere, similar to the CSBE. At high masses, however, the TNT model still shows a central `core' but also a more extended `clump'. On the other hand, the CSBET model retains the same shape but its peak column density scales inversely with mass while its extent scales in proportion to mass. Thus as mass increases, the CSBET model becomes more extended than the TNT model, while the CSBET peak drops below that of the TNT model at the same radius. The CSBEP model however doesn't show any important `core' component. It retains the same shape on the entire mass range as the CSBET does but the column density peak value is independent of the mass enclosed and is set by the external pressure.
\subsubsection{Star formation efficiency and Resolution}
\par The star formation efficiency (SFE) parameter $\epsilon$ is defined as the ratio between the star final mass $M_*$ and the initial total core mass M where the star comes from.
\begin{equation}
M_*=\epsilon M
\label{eq:SFE}
\end{equation}
\par We used  star formation efficiency values of 1.0 and 0.3 to determine the parent core mass for every star. The value of 1.0 was motivated by its use as a reference, whereas 0.3 was chosen to match the SFE value derived in the study of \cite{Alves2007} using dust extinction maps.
\par Once the parent core mass was obtained, the column density function was computed on a spatial grid of step 0.001 pc  for each YSO and for each different core model separately. This step is small enough to allow a good representation of the column density profiles (see Figure \ref{fig:model_profile}). The individual column density profiles of the progenitor cores within each stellar group were then summed together, to give the global column density map necessary to produce the stars seen today. Finally, the maps were smoothed to reproduce observational resolution due to submillimeter single-dish observations \citep[e.g., the bolometer survey in Perseus of][]{Enoch2006} or star count extinction \citep[e.g., ][whose dust extinction maps were used in \cite{Alves2007}]{Lombardi2006} with a Gaussian kernel of FWHM 0.5 arcmin and 1.0 arcmin.
\par The resulting maps can be compared directly with dust extinction maps since the extinction depends primarily on the column density. In contrast, dust emission maps are also sensitive to dust temperature; and molecular line maps depend on molecular abundance, velocity and line excitation.
\subsection{Identifying observable cores}
\par We derive the core mass function that would be observed for each group using a standard clumpfinding algorithm. We used the two-dimensional version of the automated algorithm clumpfind (clfind2d, \cite{Williams1994}), the same one as used in the Pipe nebula study of \cite{Alves2007} and \cite{Rathborne2009}. The algorithm works by first contouring the data at certain levels set by the user. Then it searches for peaks of column density which locate the cores. Afterwards the algorithm follows these peaks down to lower column density values, until it reaches the threshold set by the user. It does not thus assume any core profile as does for instance the Gaussclumps algorithm \citep{Stutzki1990} by assuming a Gaussian shape.
\par In more detail, at each iteration, the algorithm finds all the contiguous pixels which have a value between the current level and the next down one. It then assigns them to a pre-existing core or a new one, depending if they are connected or isolated from any previously identified core. In case of blended column density features, a `friends-of-friends' algorithm is used to distribute the pixels between several identified clumps. Eventually, at the final level, a core has to be greater than a certain number of pixels, if not, it is rejected. For any further details, see \cite{Williams1994}. 
\par Our choice of clfind2d parameters was based on \cite{Rathborne2009}, who in turn relied on a clfind2d performance study by \cite{Kainulainen2009}. \cite{Rathborne2009} improved on the previous list of cores compiled by \cite{Alves2007}, using contours of 2x 1.2 mag with a lowest threshold of 1.2 mag. \cite{Alves2007} used clfind2d on the Pipe nebula extinction map with a maximum contour level of 6.0 mag, which led then to the truncation of the extinction contouring at 6.0 mag and to the fusion of multiple well-separated extinction peaks into single-extinction features. Thus in our study, as in \cite{Rathborne2009}, the threshold was set to 1.2 $\times 10^{21}$cm$^{-2}$ and the difference between two levels at 2 $\times$ 1.2 $\times 10^{21}$cm$^{-2}$ (we used the conversion 1 mag = $10^{21}$cm$^{-2}$). This parameter choice was motivated by the closer comparison to \cite{Rathborne2009} that it offers.
\par One of the most important results of the simulations of \cite{Kainulainen2009} is that the degree of crowding within a molecular cloud can significantly effect both the measured core parameters and the derived CMF in a more important way than the parameters selected for the core extraction algorithm do. Therefore we separated two cases in our study, isolated and blended groups, according to the criterion used in the simulations of \cite{Kainulainen2009}. According to this criterion, a group is isolated if it displays a value of the ratio f of mean separation to mean diameter above 1.0. By separation we refer to the distance measured from the peak position of a core to the peak position of its nearest neighbor. The separation and the diameter values of the cores are defined from clfind2d output (position of the peak and surface for each core). For f $\leq$ 1, \cite{Kainulainen2009} showed that the mass determination of individual cores is very uncertain and that the derived CMF may not represent the underlying mass function. \cite{Kainulainen2009} used, for the cores, gaussian profiles, random positions and masses, whereas we are using here physical profiles, observed positions and masses of stellar groups whose most massive stars tend to be in more crowded regions \citep{Kirk2010}. These differences could influence strongly the measured CMF and increase the value of f under which the mass determination becomes uncertain; it raises also the question of the quality of the determination of the core profile. This ratio f is approximately 2.0 in the Pipe Nebula, as it is a relatively quiescent and sparse region. To avoid confusion, we denote f as the `crowding ratio' from now on.
\par In \cite{Rathborne2009}, the larger-scale background was subtracted prior to core identification. Since there is no background added to our maps, and the column density is inferred only from the present-day stars, there is no need to first subtract any background. Without background, it is true that the simulated maps departed from the observed dust extinction ones. Nonetheless, it allows us to skip the background removal step and to study more precisely the effects of the clumpfinding algorithm on the actual matter that will eventually accrete on the protostar. Finally, it is noteworthy that any additional background will increase the blending of a given region.
\section{Results}
\label{sec:Results}
\subsection{Column density maps}
\par Figures \ref{fig:isolated_blended_comparison}, \ref{fig:parameter_comparison}, \ref{fig:model_comparison_03} and \ref{fig:model_comparison_1} display the column density maps of representative groups for each region (Taurus 5 or L1529 for Taurus, IC 348 1 or IC 348-Main for IC 348, Lupus 3 1 or Lupus 3-Main for Lupus 3, ChaI 2 or ChaI-South for Chameleon I). In Figure \ref{fig:isolated_blended_comparison}, the three main steps followed in this study are summarized : starting from a YSO group (left panel), we simulate an initial starless column density map (middle panel), from which the clumpfind algorithm clfind2d isolated cores (right panel).
\par These column density maps show two distinct structural components and are then qualitatively similar to dust extinction maps, such as the maps in \cite{Kirk2006}, where large-scale structure is seen in extinction maps within which small-scale features seen in thermal emission are embedded. The simulated column density maps similarly show a `clump' component, which has a spatial extent typically around 0.3 pc (but in an extreme case like IC 348 1, it can reach 1 pc) and which originates from the external part of several summed column density profiles. On top of the clump component appear several  `core' components, which have a spatial extent less than 0.1 pc, matching the central part of a single core column density profile. This description is particularly accurate when the initial SFE is set to 0.3, which is currently the favoured SFE value.
\par Regarding clumpfind output, one can easily see by looking at the maps in Figures \ref{fig:isolated_blended_comparison} through \ref{fig:model_comparison_1} that cores are undercounted by clumpfind, although the total mass within one region is recovered. As expected, this effect is much stronger in crowded regions. Figure \ref{fig:isolated_blended_comparison} allows a quick comparison between two extreme cases : a very isolated one (Taurus 5) and a very crowded one (IC 348 1). Whereas Taurus 5 has a nearly one-to-one relationship between progenitor cores and stars, the crowded IC 348 1 shows a much higher number of stars per clumpfind core. From now on we will call the number of stars per core the fragmentation ratio F. In IC 348 1, clfind2d is less effective in identifying structures: as can be seen in Figure \ref{fig:isolated_blended_comparison}, the cores have highly irregular shapes and appear unphysical.
\par Figure \ref{fig:parameter_comparison} shows the changes in the column density map of Lupus 3 1 for the CSBET model when the values of resolution and initial SFE change. It shows qualitatively that fewer cores are found with a decrease in initial SFE and/or poorer resolution. A poorer resolution has a stronger effect however than a decrease in initial SFE, as the number of identified cores always decreases significantly with the poorer resolution.
\par Figures \ref{fig:model_comparison_03} and \ref{fig:model_comparison_1} show side by side the column density maps derived for each of the three models for the Cha I - South or Cha I 2 group, assuming a initial SFE of 0.3 (Fig. \ref{fig:model_comparison_03}) and 1.0 (Fig. \ref{fig:model_comparison_1}), and a resolution of 1.0' for both. The maps show a larger difference when the initial SFE is set to 0.3 and when the radii of massive cores are bigger. The difference in blending of sources between the panels can be explained by the relation between the enclosed core mass M and the core radius R. The CSBET model yields core radii that increase proportionally to the total enclosed mass, much more rapidly than for the CSBEP model, where R $\propto \sqrt{M}$ and the TNT model, where R $\propto$ M$^{0.73}$ (see Appendix \ref{appendix:models}, in exact form M $\propto (AR + \frac{3BR^{7/3}}{7})$, but the fit R $\propto$ M$^{0.73}$ is excellent for the present YSOs masses.). For massive stars, the CSBET model yields the widest column density profiles, which in turn causes a larger degree of blending of cores.
\subsection{Derived core properties}
\label{sec:core_properties}
\par Several properties can be derived directly from clumpfind output: the total number of cores N$_{C,tot}$, the mass of each core M$_C$, the total mass of these cores M$_{C,tot}$, the radius of each core R and the crowding ratio f describing whether the cores appear isolated (f$>$1) or blended (f$<$1). The core radius R is derived from the core surface returned by clfind2d by assuming a circular projected shape. By counting the number of YSOs one core spatially encloses, we derive the average number of stars per core, which we denote as the `fragmentation ratio' F. We fit a power-law to the function M$_C$(R) and derive the slope $\beta$ in the expression M$_C$$\propto$R$^{\beta}$. These bulk properties describing each group are given in Table 1.
\subsubsection{Multiple stars per core}
\par Before comparing the initial CMFs and the derived CMFs, it is interesting to calculate the average number of stars found in each core. Table \ref{tab:globaloutput} shows that even in isolated regions, the ratio F is never strictly equal to one and increases as the ratio f decreases, which confirms the legitimacy of the crowding ratio criterion f. Figure \ref{fig:frag_crow} shows the average value of the ratio F versus the value of the ratio f, where each point plotted is averaged over all three models and both resolutions for the same group. In addition to the spatial crowding, the number of cores identified also depends on the peak column density. A low resolution or a flatter column density profile (as in the CSBEP model) yields fewer cores identified.  
\par
\begin{landscape}
\begin{figure}[H]
	\centering
		\plotone{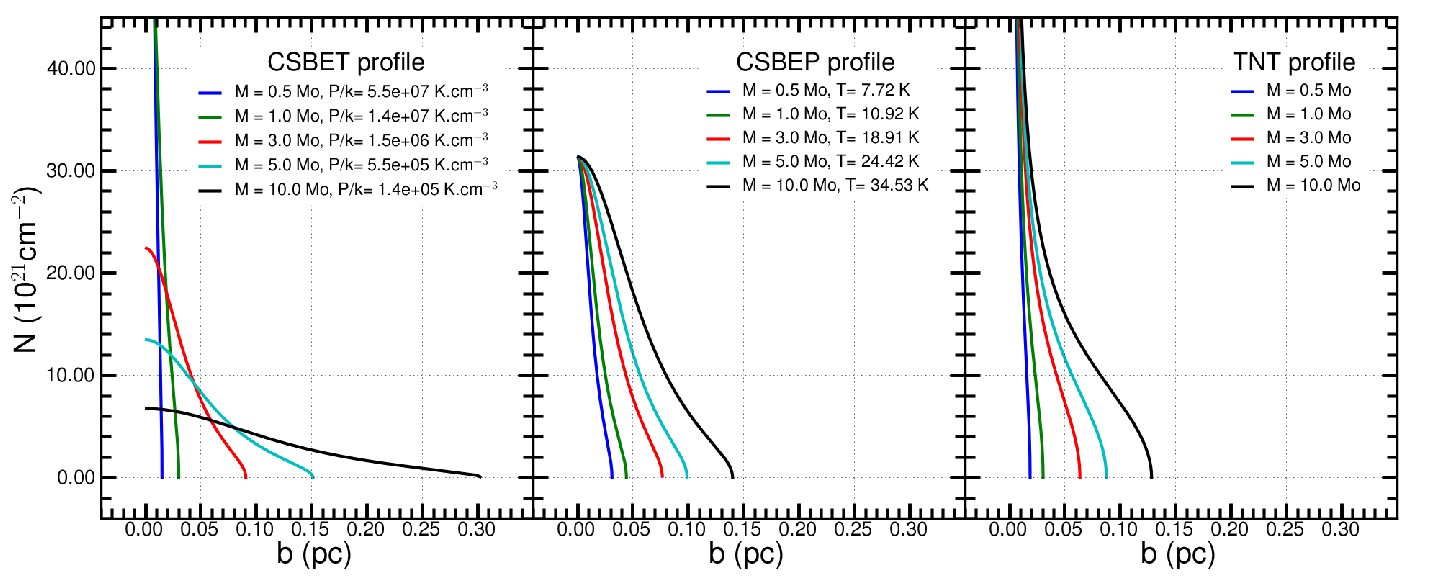}
\caption{Column density profiles enclosing 0.5, 1, 3, 5 and 10 $M_\odot$ versus the projected radius b (pc). Each plot displays a different model : \textbf{Left} : a critically stable Bonnor-Ebert sphere at temperature 16 K (CSBET), \textbf{Middle} : a critically stable Bonnor-Ebert sphere at external pressure P/k = 3.0x$10^{6}$ K.cm$^{-3}$ (CSBEP) and \textbf{Right} : a Thermal-Non thermal model (TNT).}
	\label{fig:model_profile}
\end{figure}
\end{landscape}

\begin{figure}[H]
		\plotone{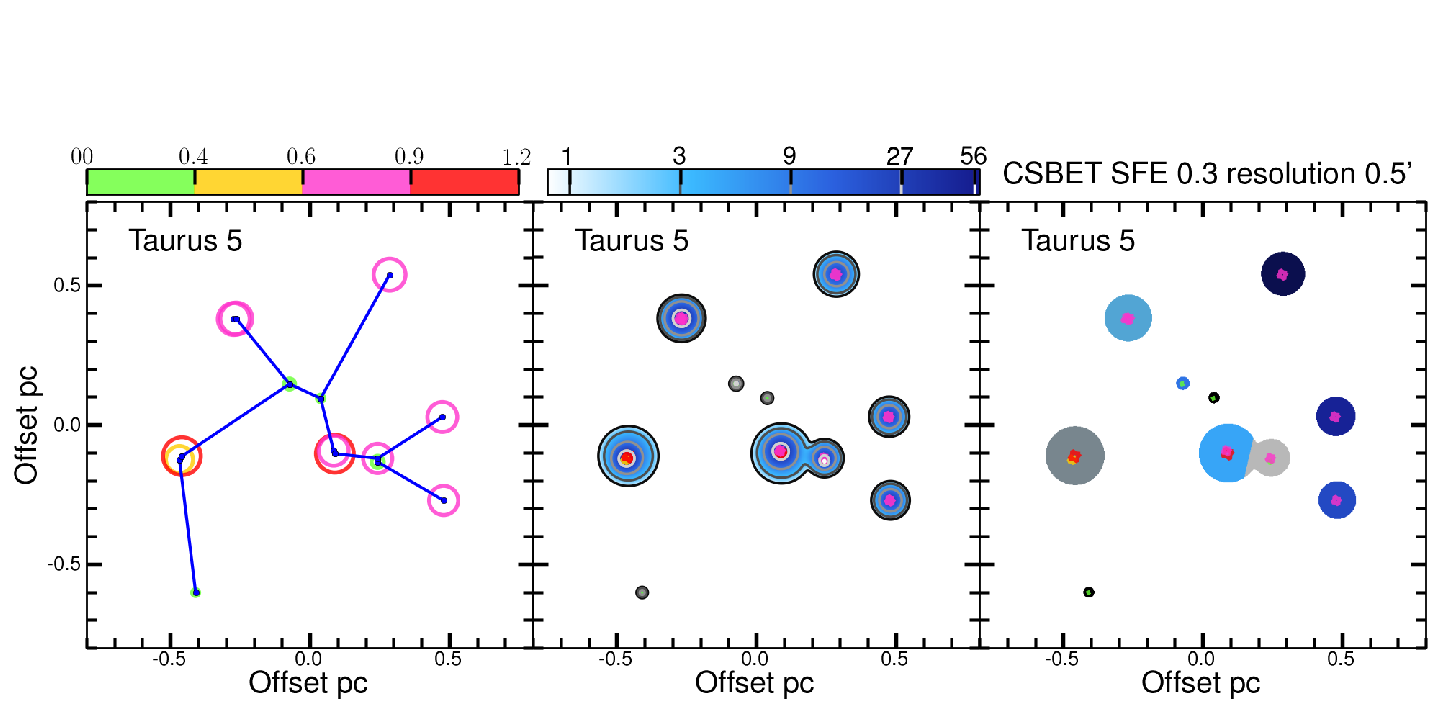}
		\vfill       \plotone{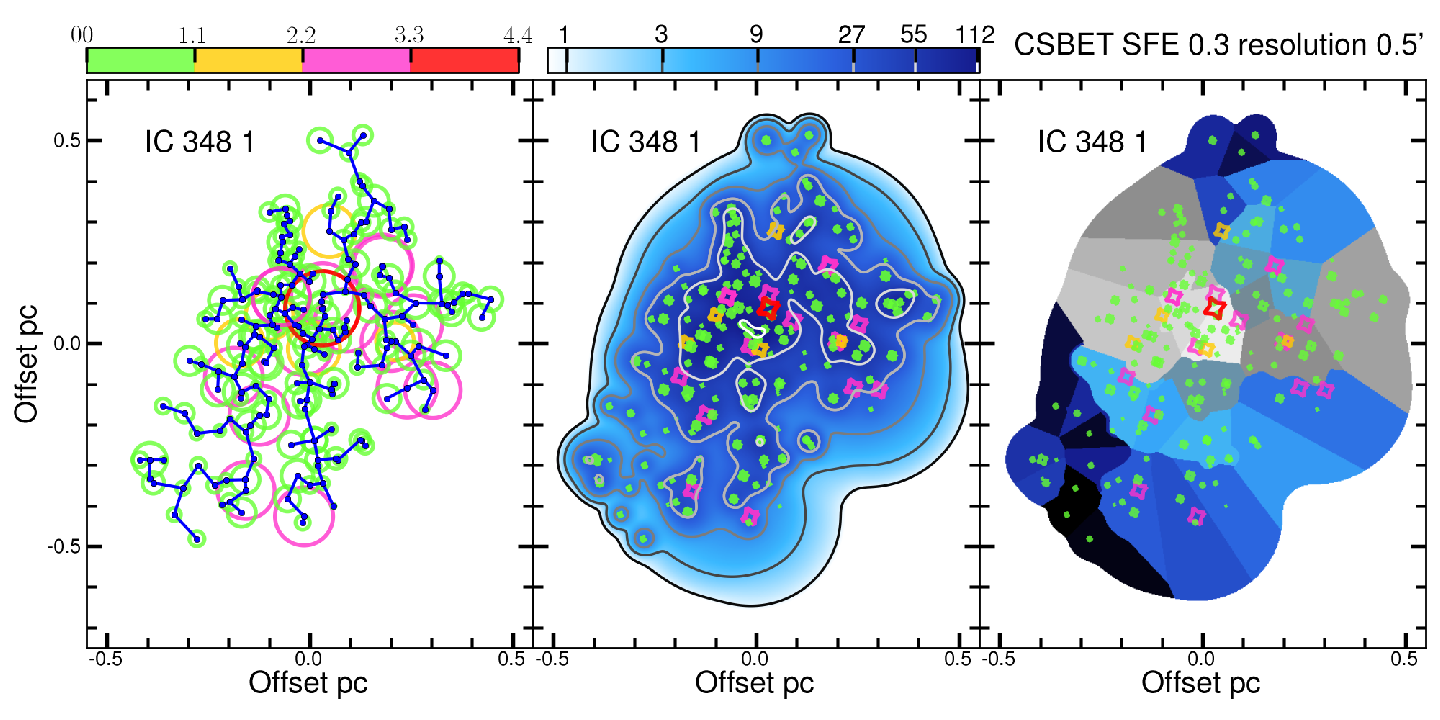}
\caption{An illustration of isolated and clustered initial condensations and the effect of blending on core identification. The top row corresponds to Taurus 5 and the bottom one to IC 348 1, for the CSBET model using a SFE of 0.3 and a blurring of 0.5'. \textbf{Left} : The initial YSO distribution is shown with circles of size increasing linearly with YSO mass and colored according to the colorbar shown on left (in M$_\odot$). The initial YSO distribution is also displayed in the middle and right figures but with markers of smaller size. \textbf{Middle} : The simulated column density map for model CSBET with SFE=0.3 and resolution of 0.5'. The blue colorbar indicates column density in units of $10^{21} cm^{-2}$. \textbf{ Right} : The resulting cores identified using clfind2d. Each core's area is shown in a different color.}
\label{fig:isolated_blended_comparison}	
\end{figure}

\begin{figure}[H]
		\plotone{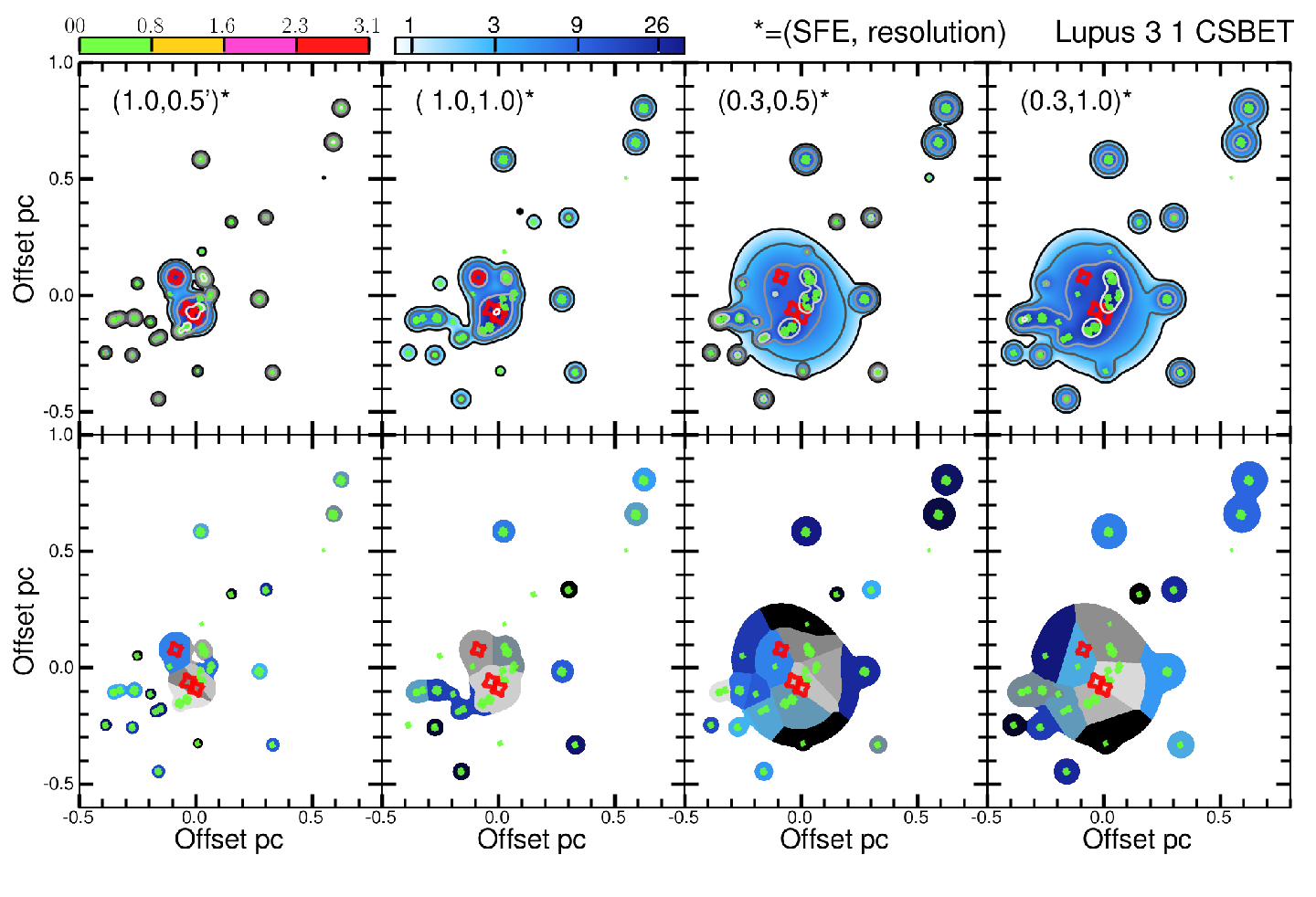} 
\caption{An illustration of how the parameters (SFE and resolution) effects the Lupus 3 1 group for the CSBET model. Along each row, each map corresponds from left to right to (SFE = 1.0, resolution = 0.5'), (SFE = 1.0, resolution = 1.0'), (SFE = 0.3, resolution = 0.5') and (SFE = 0.3, resolution = 1.0'). On each map, the initial YSO distribution is shown with marker of size increasing with YSO mass and colored according to the colorbar shown on left (in M$_\odot$). \textbf{Top} : The simulated column density map. The blue colorbar indicates the column density in units of $10^{21}$cm$^{-2}$. \textbf{Bottom}: The resulting cores identified using clfind2d. Each core's area is shown in a different color.}
\label{fig:parameter_comparison}	
\end{figure}

\begin{figure}[H]
		\plotone{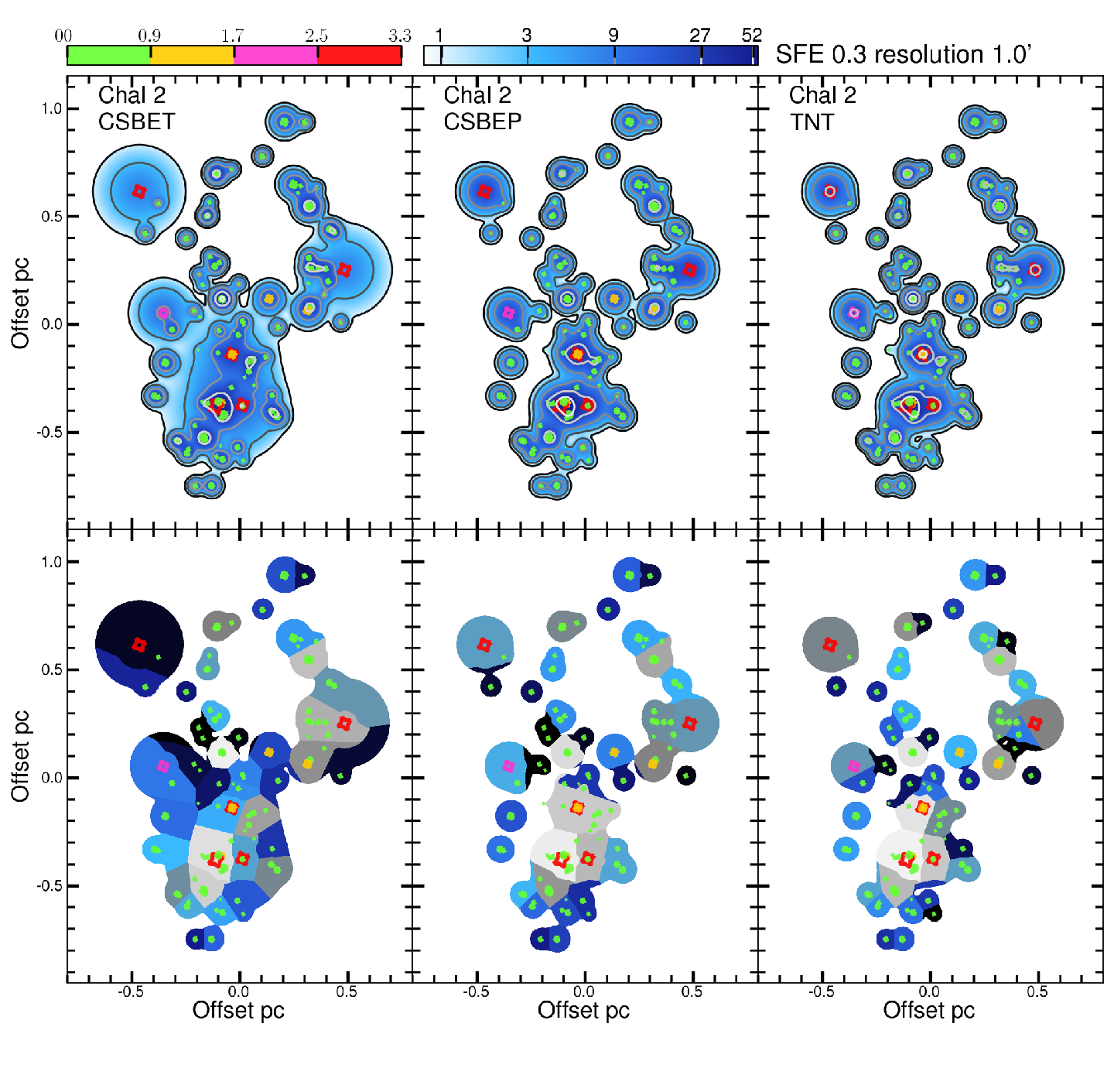} 
\caption{An illustration of the effect of using different core models in the ChaI 2 group for a SFE of 0.3 and resolution of 1.0'. Along each row, each map corresponds from left to right to the CSBET, CSBEP and TNT models. On each map, the initial YSO distribution is shown with marker of size increasing with YSO mass and colored according to the colorbar shown on left (in M$_\odot$). \textbf{Top} : The simulated column density map. The blue colorbar indicates column density in units of $10^{21}$cm$^{-2}$. \textbf{Bottom}: The resulting cores identified using clfind2d. Each core's area is shown in a different color.}
\label{fig:model_comparison_03}	
\end{figure}

\begin{figure}[H]
		\plotone{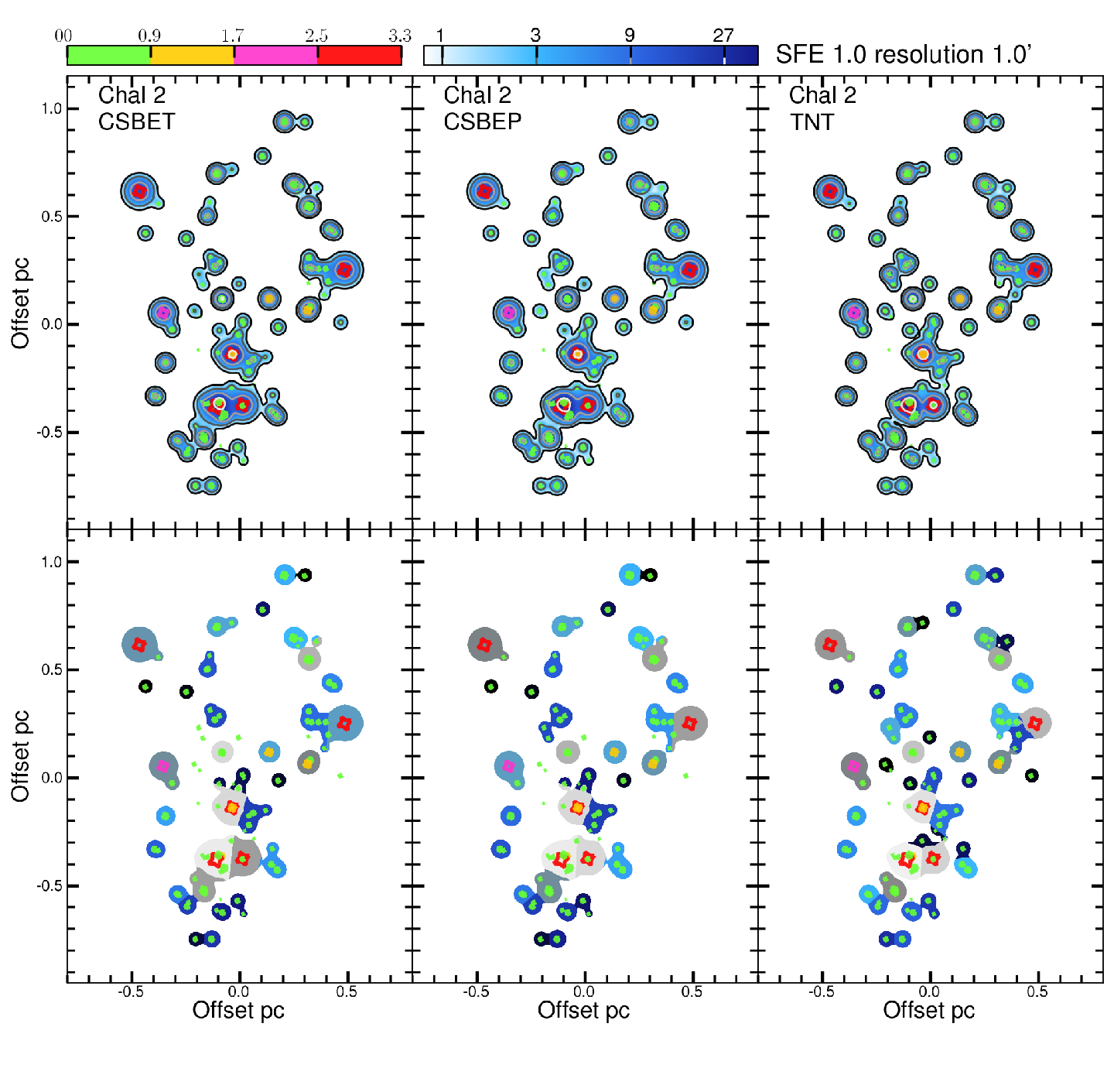} 
\caption{An illustration of the effect of using different core model in Cha I 2 group for a SFE of 1.0 and resolution of 1.0'. Along each row, each map corresponds respectively from left to right to CSBET, CSBEP and TNT model. On each map, the initial YSO distribution is shown with marker of size increasing with YSO mass and colored according to the colorbar shown on left (in M$_\odot$). \textbf{Top} : The simulated column density map. The blue colorbar indicates column density in units of $10^{21}$cm$^{-2}$. \textbf{Bottom}: The resulting cores identified using clfind2d. Each core's area is shown in a different color.}
\label{fig:model_comparison_1}	
\end{figure}

\begin{figure}[H]
\centering
		\plotone{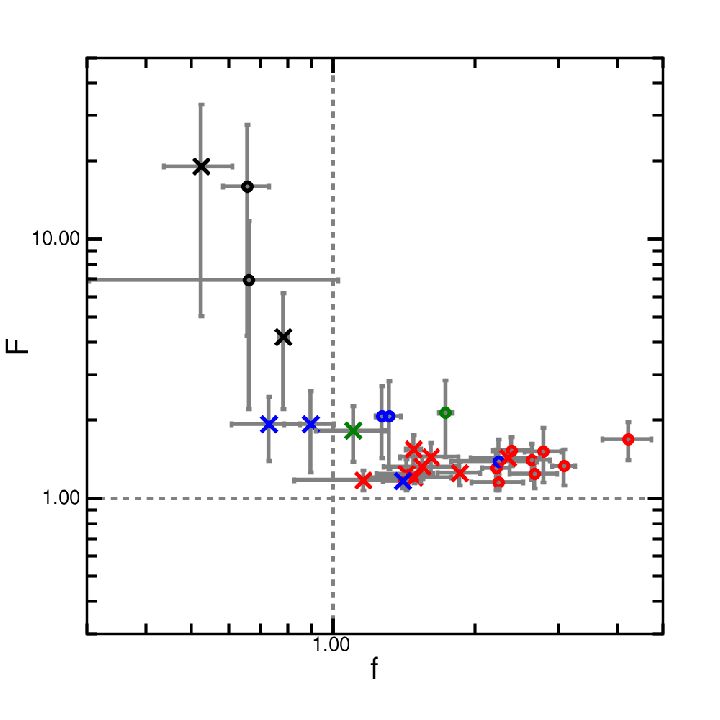} 
\caption{\small The fragmentation ratio F versus the crowding ratio f. Each color represents a region (Red : Taurus, Blue : Chamaeleon, Green : Lupus 3 and Black : IC 348) and each group is represented by a circle for the SFE value of 1 and by a cross for the SFE value of 0.3. Each point is positioned at the average value over the three models for the two resolutions while the bars at each point represent the minimum and maximum values for the crowding ratio (horizontally) and the fragmentation ratio (vertically) over the three models and two resolutions. Strong crowding (f$<<$1) results in a large number of input cores merging in the map, i.e., a high fragmentation ratio F.}
\label{fig:frag_crow}	
\end{figure}

 For a given set of parameters and group, the CSBEP model almost always leads to the highest fragmentation ratio value of the three models. This may seem surprising as the CSBEP model has the smallest radii for massive cores and as such, the least blended maps for a initial SFE of 0.3. The high fragmentation can be explained by the fact that the peaks of CSBEP profiles are the broadest -- the CSBEP model does not lead to a `core' component which is as important as the CSBET model for low masses cores or the TNT model for any core (see Figure \ref{fig:model_profile}).
\par The CSBET model, however, has the highest fragmentation when the initial SFE is 0.3 and thus the core masses are much bigger. The CSBET column density peaks are then smaller and broader (see Figure \ref{fig:model_profile}). On the contrary, the `core' component of the TNT model remains narrow in both low and high mass ranges. The broadness of the column density peaks has a larger effect on the fragmentation than the initial SFE.
\par Thus the blending of sources leads itself to undercounting, all the more important when the column density profiles are not spiky enough to lead to a good core identification. Such undercounting appears even in groups where the blending is very low (e.g., Taurus region).
\subsubsection{Incomplete mass recovery}
\par Now that it appears that the common observational CMF derivation using clfind2d is not reliable for recovering initial cores in blended regions, we want to know how the core mass found by clfind2d relates to the input mass of the YSOs within the identified core. Similarly we seek the resulting relation on the relation between the initial CMF and the derived CMF. The initial CMF corresponds by a one-to-one relationship to the local YSO distribution. We know already that, as clfind2d uses a threshold, there will always be some material that won't be assigned to any core. Hence the total mass of cores will be less than the total mass of gas in the column density map. A poorer resolution strengthens this effect as the maps are smoother.
\par For each derived core, we compared in Figure \ref{fig:mass_loss} the mass assigned by clumpfind and the mass it should have, given the YSOs it encloses. In isolated regions (Taurus and to a lesser extent Chamaeleon I and Lupus 3), for the CSBET and CSBEP models, the most massive derived cores are up to 70 \% less massive than they should be, while the least massive derived cores lack around 30 \% of their mass. In the same regions, the least massive TNT derived cores get assigned \textit{more} material than they should get by a factor of 30 \% too. The most massive TNT derived cores lack up to 50 \% of their mass. The differences between the models can be explained by the differences in column density profiles : the TNT model always gives a strong core component, whereas the CSBEP model shows no strong core component and the CSBET model shows only a strong core component at low masses. In crowded regions (IC 348 or Chamaeleon I and Lupus 3 when the initial SFE is set to 0.3), the mass from the most massive initial cores tends to be spread out sufficiently for the massive cores found by clfind2d to miss 50-75 \% of their input mass, with some of that material being added to the surrounding low and intermediate mass cores.
\par The comparison between the initial CMFs and the derived CMFs is thus complicated by these effects. If low-mass derived cores are less (respectively more) massive than they should, the derived CMFs are broadened (respectively narrowed) in this mass range. When the blending is important, the masses assigned to derived cores are not related to the YSOs they enclose, particularly for the CSBET model which yields the most blended maps.
\subsubsection{Core radii}
\par
\begin{figure}[H]
		\plotone{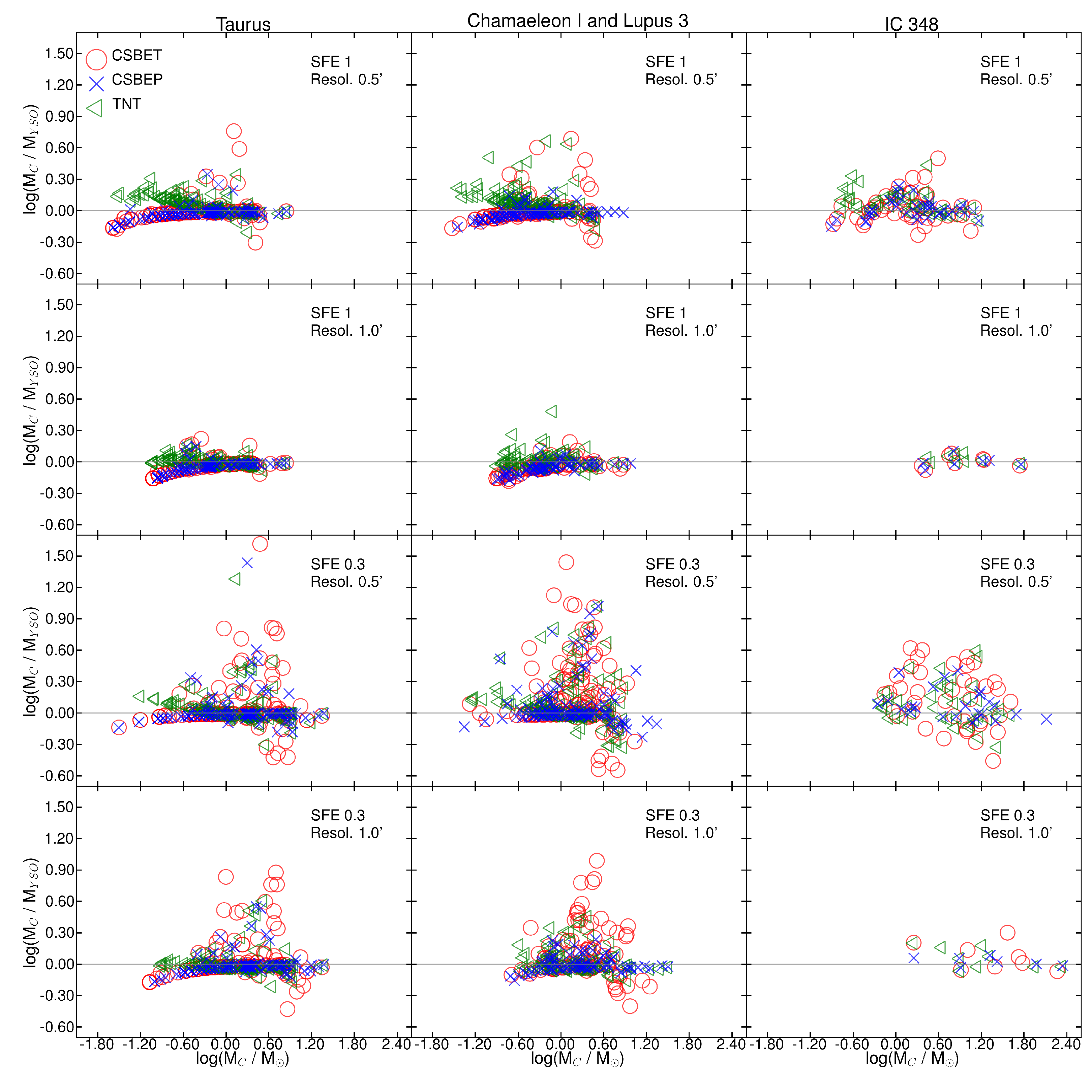} 
\caption{\small Each column shows for one region the log of the ratio of the mass derived by clumpfind for a core (M$_C$) and the mass the same core should have to produce the YSOs it encloses (M$_{YSO}$) versus  log(M$_C$). ChaI and Lupus 3 were regrouped as their properties of crowding are much alike. Along each column, each map corresponds respectively to a different choice of parameters (SFE,resolution) and displays the results for each model. CSBET and CSBEP shows similar trends which are very different from the TNT results. These trends seem to disappear as the crowding increases and the SFE decreases.}
\label{fig:mass_loss}	
\end{figure}

 In the two last sections, we have shown that the derived CMF depends on the input model for relatively isolated regions, as the fragmentation ratio F and the way a core mass is related to the masses of the YSOs it encloses change with the input model (see Figure \ref{fig:mass_loss} column 1 and 2, two top panels corresponding to a resolution of 0.5'). It is then reasonable to think that one can identify the model used to create a map, at least in isolated regions. The best way to achieve this is to study the relation between the core radius R and the core mass M$_C$. Each input model is defined by a particular power law relationship (CSBET : M$_C$~$\propto~R$, CSBEP : M$_C$~$\propto~R^{2}$ and TNT :  M$_C$$\propto$ R$^{1.4}$ for a SFE of 1 and given the YSO distributions and  M$_C$$\propto$ R$^{1.6}$ for a SFE of 0.3 and given the YSO distributions, see Appendix \ref{appendix:models}).
\par Figure \ref{fig:radius_mass} shows derived core radius R versus derived core mass M$_C$, each column corresponding to a region and each line to a set of initial SFE and resolution.
\par The power law M$_C \propto R^{\beta}$ was fit by using a least squares method. It appears that the derived power law depends on the degree of crowding, the initial SFE and the resolution rather than the initial model power law. Strong blending yields a higher $\beta$ exponent, since low mass derived cores get assigned nearby gas which are actually from the external part of more extended and massive nearby derived cores. More strikingly, the value $\beta$ is in most of the cases around 3, which matches a constant density profile.
\par It appears then that the blending of cores in crowded regions dominates the relation between core radius and mass; the input model relationship has little effect and one cannot discriminate among initial core models using derived core properties.
\medskip
\par We show in this section \ref{sec:core_properties} that the blending, which occurs with no noise and no additional background material, leads to a significant undercounting of cores (from a factor of 1.4 in isolated groups up to a factor of 20 in blended groups, see Figure \ref{fig:frag_crow}), whose assigned masses do not relate to the YSO they enclose in case of strong blending (see Figure \ref{fig:mass_loss}). Moreover, core extraction and mass recovery depends on the shape of core profile (see Figure \ref{fig:mass_loss}), but the relationship between the mass M and radius R of a derived core is independent of the input core profile and obeys a power-law of M$\propto$R$^3$ (see Figure \ref{fig:radius_mass}). 
\subsection{Derived core mass distributions}
\par We now regard the global scale of a group by comparing the mass function of the derived cores, the derived CMF, and the mass function of the initial cores, ie the mass function of the original YSOs of the group modulo some star formation efficiency,the initial CMF. Usually observationally derived CMFs are compared directly to the IMF as the local star distribution is not available. Here, however, we do not have to assume anything about the future star distribution.
\par Figures \ref{fig:taurushisto} through \ref{fig:ic348histo} show the original YSO and clfind2d derived cores binned mass functions for each region and each set of parameters. As discussed earlier, the number of cores identified is always smaller than the number of stars. The discrepancy becomes more severe in crowded regions (f$<$1), especially for a poor resolution and a low initial SFE. In the less blended configurations, the derived core mass distributions and the initial core mass distributions do not show a noticeable shift between their peaks. When the configuration becomes more blended, a shift between the peaks of the distributions appears, with the magnitude varying with the degree of crowding. For instance, in IC 348 for the CSBET model (Figure \ref{fig:ic348histo}, first row), the shift between the peaks of the two distributions is about a factor of ten, when the initial SFE is set to 1 and the resolution to 0.5', and increases to a factor of 30 when the initial SFE is set to 0.3 and the resolution to 1.0'.
\par The most important parameters to describe the IMF are the peak mass, where the function reaches its maximum (around 0.1 M$_\odot$ \cite[see][]{Kroupa2002,Chabrier2003}), the break mass, from which the scale free power law behavior seems to hold for the high mass range (around 1 M$_\odot$ \cite[see][]{Kroupa2002,Chabrier2003}) and the slope of the high mass tail power law (around 2.3 \cite[see][]{Salpeter1955,Kroupa2002,Chabrier2003}). To compare an observed CMF and the IMF, it is usual to compare these parameters for both functions, the star formation efficiency being defined as the ratio (see equation \ref{eq:SFE}) between the peak masses or the break masses of the two functions. The SFE derived from the comparaison between the derived CMF and the initial YSO distribution will be called `derived' SFE. Here we reproduce this comparison for the relationship between the derived CMF and the local star mass distribution, matching the initial CMF with the initial SFE shift.
\par To make a power law fit of the Differential Core Mass Function (DCMF, $dN/dM=p(M) \propto M^{-\alpha}$), we used the maximum likehood estimate (known as MLE, see \cite{Clauset2007} for further details, and also \cite{Pineda2009}). This method avoids both the problems raised by analyzing the cumulative function and binning the data. As pointed out in \cite{Munoz2007} and \cite{Rosolowsky2005}, the upper mass limit can induce high curvature at the high mass end in the cumulative function shape. Moreover cumulative functions can show curvature leading to a multiple power law fit, even in the case where the underlying CMF can be characterized by a single power law  \citep[see][]{Li2007}. On the other hand, binning the data induces most likely information loss \citep[see][]{Rosolowsky2005}. The MLE consists of fitting the following function :
\begin{equation}
\frac{dN}{dM} = N_{cl}\frac{\alpha - 1}{M_{Break}}\left(\frac{M}{M_{Break}}\right)^{-\alpha}
\end{equation}
 where $M_{Break}$ is the minimum mass value at which power-law behavior holds. $N_{cl}$ is the number of cores more massive than $M_{Break}$ and $\alpha$ is the power-law exponent of the distribution. The MLE gives an estimate of the exponent and the approximated standard error $\sigma_\alpha$ on it :
\begin{equation}
\alpha = 1 + N_{cl}\left[\sum^{N_{cl}}_{i = 1}ln\left(\frac{M_i}{M_{Break}}\right)\right]^{-1}
\end{equation}
\begin{equation}
\sigma_\alpha = \frac{\alpha - 1}{\sqrt{N_{cl}}}
\end{equation}
\par The lower cut-off for the power-law region, $M_{Break}$, was determined by a Kolmogorov – Smirnov test. All of these steps were performed using an algorithm based on plfit.py, a python implementation of Adam Ginsburg based on the general algorithm of \cite{Clauset2007}, (see also online \url{http://tuvalu.santafe.edu/~aaronc/powerlaws/} and for the python implementation \url{agpy.googlecode.com}.).
\par Figure \ref{fig:fit_quality} shows the fit obtained using the MLE method for Cha I 2 for the TNT model, initial SFE 0.3 and resolution 1.0'. It appears that the slope obtained for the core mass function is shallower than the one obtained for the star mass function. In the MLE method, a finite size bias can be present when the number of cores of mass above M$_{Break}$ is under 50 \citep{Clauset2007}. As some groups don't have enough cores to allow a good determination of $M_{Break}$ and $\alpha$, we divided the groups into two categories for each initial SFE and core model: blended groups, which have a crowding ratio f $\leq$ 1 for at least one resolution value and isolated groups for the remaining ones. The blended groups are listed in Table \ref{tab:repartition} for each initial SFE and core model. In the isolated groups, there are always more than 50 cores identified. In the blended groups, however, there are fewer than 50 cores above M$_{Break}$ when the resolution is 1', for every model and initial SFE. There the fragmentation ratio F is high and the total number of cores very low.
\par The results of the power-law MLE fit are displayed on Figure \ref{fig:sfe}. For each input SFE (1.0 and 0.3), the derived SFE values are shown in the upper plot and the $\alpha$ values are shown in the lower plot. The derived SFE is estimated in two ways : SFE$_{Peak}$ is the ratio of the peak masses of the stellar and the derived core distributions and SFE$_{Break}$ is the ratio of the break masses of the stellar and the derived core distributions. M$_{Peak}$ is estimated as the position of the maximum in the mass distributions using a Gaussian kernel because, as this does not require discrete data bins, it more faithfully reproduces the detailed structure of the CMF compared to a binned mass function \citep{Silverman1986}. M$_{Break}$ and $\alpha$ are from the MLE power law fit.
\par The fit results on Figure \ref{fig:sfe} show how the peak and slope of the derived CMF compare to the peak and slope of the distribution of stars. For a one-to-one relation between derived cores and stars, with constant SFE, the peak and break masses should follow M$_{peak}$(stars) = (SFE$_{peak}$)$\times$M$_{peak}$(cores), M$_{break}$(stars) = (SFE$_{break}$)$\times$M$_{break}$(cores)  and the slopes should be identical. Figures \ref{fig:taurushisto} - \ref{fig:ic348histo}, however, show that crowding and poor resolution generally results in a one-to-many relation between cores and stars.
\par In isolated regions, the initial SFE is recovered by both SFE$_{Peak}$ and SFE$_{Break}$ as the difference between the initial SFE and the derived SFE$_{Peak}$ and SFE$_{Break}$ are only a few percent, with the largest difference at a resolution of 1.0'. The mass at the peak of the derived CMF exceeds the mass at the peak of the initial CMF by a mean factor 1.0. In blended regions, however, SFE$_{Peak}$ and SFE$_{Break}$ are not similar either to the input values or one another. For a resolution of 0.5', the SFE$_{Peak}$ value is around 0.1 and the SFE$_{Break}$ value is around 0.15/0.25  for the three models and both initial SFE input. For a resolution of 1.0', all the models for a initial SFE of 1.0 have a SFE$_{Peak}$ value around 0.03 and a SFE$_{Break}$ around 0.06, while all the models for a initial SFE of 0.3 have both SFE$_{Peak}$ and SFE$_{Break}$ of around 0.1. The mass at the peak of the derived CMF exceeds the mass at the peak of the initial CMF by a mean factor 12.1.
\par Regarding the slopes, all the derived slopes, both for isolated and blended groups, are similar to the value of 2.35 from \cite{Salpeter1955} within errors. In comparison to the slope derived in the local star distribution, the slopes are slightly better recovered in the blended regions. In blended regions, the stellar slope value is around 2.4 and the derived core slopes range from 2 to 2.5. In isolated regions, however, the stellar slope value is around 3 and the derived core slopes range from 2.5 to 2.7.
\par To discuss directly the links between the different mass ranges and their properties, the resulting cumulative mass functions E(M), the fraction of cores or stars with mass greater than M, are shown in Figure \ref{fig:cumulo_I} for the isolated groups and in Figure \ref{fig:cumulo_B} for the blended groups. Each row corresponds to a different input model and each column to a different set of initial SFE and resolution. Each panel shows the cumulative functions with the power-law MLE fit for the derived cores, the local star distribution shifted by the initial SFE value and the Chabrier IMF \citep[see][]{Chabrier2003}. The shift between the local star distributions and the Chabrier IMF shows that variations from region to region regarding the IMF must be considered when comparing IMF and CMF, as stated in \cite{Swift2008}.
\par In isolated regions, the fragmentation ratio is very close to one, which allows a good recovery of the initial SFE by both methods (SFE$_{Peak}$ and SFE$_{Break}$). Since the fragmentation ratio is not exactly one, some derived cores are actually several initial ones blended together. As the more massive initial cores have a more extended profile and are almost always positioned in or near the most clustered part of the group \citep[see][]{Kirk2010}, the blending of initial cores happens more often for massive cores. Therefore, the positions of M$_{Peak}$ and M$_{Break}$ for the derived core distribution are very similar to the ones for the star distribution since only the most massive initial cores are concerned by the blending in those isolated regions. The slope, however, is very sensitive to the high mass tail and the derived CMF slopes are shallower than the initial stellar distribution ones. In Figure \ref{fig:cumulo_I}, the comparison of the star and derived core cumulative mass functions shows that, for a resolution of 0.5' (columns 1 and 3),the two distributions only depart in the very last part of the high mass tail, as the fragmentation ratio F is very close to one. For a resolution of 1' (columns 2 and 4), the fragmentation ratio F is less close to one and the shift between the two distributions happens at a lower mass. This does not change the derived SFE, but the slopes of the derived cores and stellar distributions differ more than in the better resolution case.
\par In blended groups, the blending of initial cores is much more important, which is a difficulty when trying to derive the SFE and the slope $\alpha$. The number of derived cores can be very small (see Figure \ref{fig:cumulo_B}, column 2, where only a few cores are identified). Even when the number of derived cores is sufficient to avoid the MLE fitting finite size bias, the derived SFEs are very different from the initial value (for instance, the CSBET model with an initial SFE of 0.3 has a SFE$_{Peak}$ of 0.10 and a SFE$_{Break}$ of 0.25 for a resolution of 0.5', and 0.09 and 0.12 respectively for a resolution of 1.0'). As can be seen in Figure \ref{fig:cumulo_B} in all panels, the star and derived core distributions are separated by a shift that is caused by a fragmentation ratio F $>$1. The shift is wider when the resolution is poor -- compare columns 1 and 3 with a resolution of 0.5' and columns 2 and 4 with a resolution of 1'. This effect is directly translated in the derived SFE values which can drop up to 70\% when the resolution changes from 0.5' to 1.0'. As in isolated groups, the derived slopes are shallower than the initial ones: as for the isolated groups, several initial cores are blended together producing only one more massive derived core. This happens preferentially for the most massive initial cores as they are larger and tend to be in a more cluster part of the group.

\par The blending, caused by a spatial crowding and a low resolution, seems then to be the crucial parameter when deriving the SFE value. It is well recovered in isolated groups (within a few percents of the input value, see Figure \ref{fig:cumulo_I} and \ref{fig:sfe}), but not in blended groups, where the SFE is underestimated and independent of the input value (around 0.15 for a resolution of 0.5', around 0.05 for a resolution of 1', the peak mass of the derived CMF exceeds the peak mass of the initial CMF by a mean factor of 12.3 see Figure \ref{fig:cumulo_B} and \ref{fig:sfe}). Regarding the recovery of the slope $\alpha$, it is shallower than the initial one even in isolated regions but can always be fitted by power law whose slopes are similar to Salpeter value (2.35) within error for both isolated and blended groups (see Figure \ref{fig:cumulo_I}, \ref{fig:cumulo_B} and \ref{fig:sfe}). The comparison of derived CMFs does not distinguish the input models from one another. In addition to causing the differences between derived core masses and radii compared to the input models, the blending affects also the properties of groups on a global scale, as in no case do we obtain a quantitative recovery of the initial CMF by the derived CMF.

\section{Discussion}
\label{sec:Discussion}
\par Starting from YSO masses and positions in 4 nearby star-forming regions, we carried out simulations to estimate the initial starless core column density maps with a one-to-one relation between YSOs and cores and with different values of SFE. The maps are also smoothed with different resolution values. After running the clumpfinding algorithm clfind2d on these maps, we derived the CMF for both isolated groups and blended groups. If our procedure were `perfectly recursive', we would recover each and every stellar mass that we started with, modulo the SFE. In no case, however, do we obtain such recovery because of the blending. The blending arise from spatial crowding of cores but also from the smoothing effect of the resolution. The blending appears for all three initial core column density models considered.
\par The difficulty to recover the initial stellar distribution can not be a consequence of the choice of parameters regarding the clumpfind algorithm in a significant way. The threshold of 1.2$\times$10$^{21}$cm$^{-2}$ is low enough to describe well all regions, as it can be observed for representative groups for each region on Figures 2, 3, 4 and 5 as simulated column density maps and clumpfind outputs show similar boundaries. Contouring levels of 2$\times$ the threshold describe well the column density landscape of even the more crowded regions. Figure \ref{fig:levels} shows the contouring levels for ChaI 3 and IC 348 1 in the case of the TNT model, SFE of 0.3 and resolution of 1.0' : the difficulty of determining the peaks of different cores is clearly a consequence of the blending rather than of the contouring levels value. Using finer contouring levels could lead to a significantly better core identification only if they were of small fractions of an Av, which is much smaller than observational uncertainties.
\par In both isolated and blended groups, it is always possible to fit the high-mass tail of the recovered mass function with a power law whose slope is within 0.2-0.3 of the Salpeter value of 2.35, the errors being in average of $\pm$0.2. These good fits do not discriminate among models. The MLE fitting method requires a minimum of 50 cores above the break mass to avoid finite size bias, which was not fulfilled in blended regions when the resolution is 1.0' and there is strong blending. Despite this, the properties derived in those cases are consistent with the properties derived in blended regions when the number of cores is sufficient. The fact that we find a Salpeter power-law even in the most blended groups where most of the initial cores are merged together or when the original YSO mass function shows a steeper high mass tail, could be the consequence of the central limit theorem, as is argued in \cite{Reid2010}.
\par In isolated groups (crowding ratio f $>$1), our procedure is qualitatively sucessful, but undercounts the cores by a typical factor of 0.7. Their masses are globally recovered, as both derived SFE$_{Peak}$ and SFE$_{Break}$ are within a few percents of the initial SFE. The fraction of individual core mass recovered, however, can vary significantly. In case of poor resolution (1'), massive derived cores lack up to 70 \% of their mass. The recovered mass depends on the core density profile and resolution as clfind2d is sensitive to spiky and not extended structure. The reason for qualitative global recovery is that blending effects tend to be local, only a few cores being merged together at a time.
\par In blended groups (crowding ratio f $<$1), this procedure is qualitatively and quantitatively unsuccessful, because progenitor cores blend into much larger clumps, which cannot be parsed into their initial cores. Reasons for this are first that massive cores possess a very extended profile and are almost always positioned in a location of higher than average stellar surface density \citep[see][]{Kirk2010}, which leads to the blending of core profiles on a more global scale than in isolated groups. Second, clfind2d identifies first the peak of a core and then assigns it an extended structure. It leads then to the undercounting when individual peaks are merged into the summed column density map. As a poor resolution smooths the peaks, it worsens also the undercounting.
\par These results suggest that, if stars are represented by progenitor cores with a fixed mass ratio, the distribution of apparent core masses derived from a column density map is an unreliable estimator of the stellar masses, especially in young clusters. The blending effects, as the horizontal shift between IMF and CMF, may be easily mistaken with physical effects (low SFE, fragmentation process, different evolutionary timescales).
\par The blending issue was already addressed by \cite{Hatchell2008}. In Perseus, the mass distribution of prestellar cores shows an excess of prestellar cores at low masses. A possible explanation advanced by \cite{Hatchell2008} is the selection effect due to the blending. This effect is increased by the fact that all clumpfinding algorithms look for peaks and can not then identify a core whose peak is merged with profile of other cores. The extent of this effect, in comparison to the effect of particular evolutionary schemes, as different evolutionary timescales for different cores, remains unknown in Perseus. The difficulty in retrieving the evolutionary scheme by comparing IMF and CMF was also addressed by \cite{Swift2008}. This difficulty is increased by the impossibility of comparing an observational CMF directly to the IMF of stars that will actually form. The present results come from the comparison between the local mass star distribution and the derived CMF, i.e. the `observed' CMF, and it is not possible to distinguish different evolutionary scheme, apart from the SFE in the isolated groups.
\par The smoothed column density maps are similar to observational dust extinction maps \citep[see][]{Kirk2006}. Moreover the maps appear to meet the conclusions of \cite{Smith2009}, particularly in blended case. This study found by using smoothed particle hydrodynamic simulations of massive star-forming clumps that most of the mass accreted by the massive stars was originally distributed throughout the clump at low densities. The radius values in the maps derived here and the results in \cite{Smith2009} are also similar, as very massive cores reach radii around 0.5 pc and low-mass ones have radii less than half a tenth of parsec.
\par Observations, however, can also detect cores which will disperse before they form any stars. Thus the observed CMF is a mixture of lower-mass cores which make no star, and higher-mass cores which make more than one star \citep[see][]{Hatchell2008}. In our column density maps, we did not add any background or additional low-mass cores that will not form stars. Our maps are then less close to observations but show the best possible observable relationship between cores and stars. Any extra background or lower-mass cores which make no star will increase the local blending and strengthens our main results about the crucial role played by the blending in the derivation of any CMF.
\par Another limitation of our results is that the stars could have moved from their birth sites and then the column density maps are not representative of what the actual initial column density maps would have been. The motions of stars in the first few Myrs, however, are subvirial \citep[see][]{André2007,Offner2009}, i.e., relatively slow motions and the configuration of stars should not have changed much since their birth. Even so, net motions of stars are more outward than inward, so our approach probably underestimates actual blending.
\par The maps are simulated by adding the column density profile of every progenitor core and thus by assuming an identical formation moment for every star. A way of checking the relevancy of the simulated starless column density maps would be to do a SPH simulation from the simulated maps back to a YSO distribution and compare it to present YSO distributions. This SPH simulation should take into account different star formation moments. It would however require assumptions about how the original cores are distributed along the line of sight, as we only know the YSO positions in the plane of sky.
\par By being able to compare directly the derived CMFs to the local stellar distribution, our approach allows good estimates of the undercounting of cores due to the blending, as we start with a one-to-one relation between stars and simulated progenitor cores. The differences between real star-forming regions and the young clusters used here underestimate the effects of blending in actual clusters.
\section{Conclusions}
\par Starting from YSO masses and positions in 4 nearby star-forming regions, we carried out simulations to estimate the initial starless core column density maps. After running the clumpfinding algorithm clfind2d on these maps, we find that an derived CMF can have the right `shape' to match IMF but can nonetheless undercount the number of star-forming cores and overestimate their masses, due to blending of initial cores in crowded regions. Such blending can occur even with no additional background material, no noise, a 100\% star formation efficiency. Our results are :
\begin{itemize}
\item In no case do we obtain a quantitative recovery of the initial CMF by the derived CMF.
\item The comparison of derived CMFs does not distinguish the input models from one another.
\item The derived high-mass tails can always be fitted by power law whose slopes are similar to Salpeter value (2.35) within errors for both isolated and blended groups.
\item Even in isolated groups, derived cores are undercounted by a factor of 1.4. In blended groups, this factor can be as high as 20.
\item The initial SFE is recovered within a few percents in isolated groups, whereas it is not recovered in blended groups, where the initial SFE is underestimated and independent of the input SFE (around 0.15 for a resolution of 0.5', around 0.05 for a resolution of 1').
\item The peak mass of the derived CMF exceeds the peak mass of the initial CMF by a mean factor of 1.0 in isolated regions and by a mean factor of 12.3 in blended regions.
\item Mass recovery depends on shape of core profile.
\item The relationship between the mass M and radius R of a derived core is independent of the input cores profile and obeys a power-law of M$\propto$R$^3$.
\end{itemize}
\par These results suggest that, if stars are represented by progenitor cores with a fixed mass ratio, the distribution of apparent core masses derived from a column density map is an unreliable estimator of the stellar masses, especially in young clusters.
\par Deriving an accurate CMF in blended regions appears to be a very difficult task. Clumpfinding programs which rely on peak identification have difficulties identifying the extended part of the core profile in blended maps. Even in more isolated groups, the outer part of the cores is not well recovered and we find the relationship between derived core masses and radii is independent of the input model relationship. Further difficulties are expected to arise in real observations where an extended background is removed, as the very extended component of the core could be mistaken for the background or even be missed if it falls below the detection limit of the observations.
\par We have examined the mapping of known YSOs distributions to their modeled observable progenitor cores. The conclusions of this work should be very similar what observations would give if the local star distribution was available at the same time as the core mass function. We expect the results to underestimate the effects of the blending, as real observation must deal with background removal and likely a slightly more compact configuration of cores than their present-day locations.
\begin{acknowledgements}
We would like to thank Jaime Pineda for very interesting discussions about the clfind2d clumpfinding algorithm, Michael Reid for sending us his paper before publication and discussions about observational bias in deriving CMF and Patrick Hennebelle for his comments. We are also thankful to the Smithsonian Astrophysical Observatory for partial support of the visit of Manon Michel.
\end{acknowledgements}
\bibliographystyle{apj}  

\newpage
\begin{figure}[H]
\centering
		\plotone{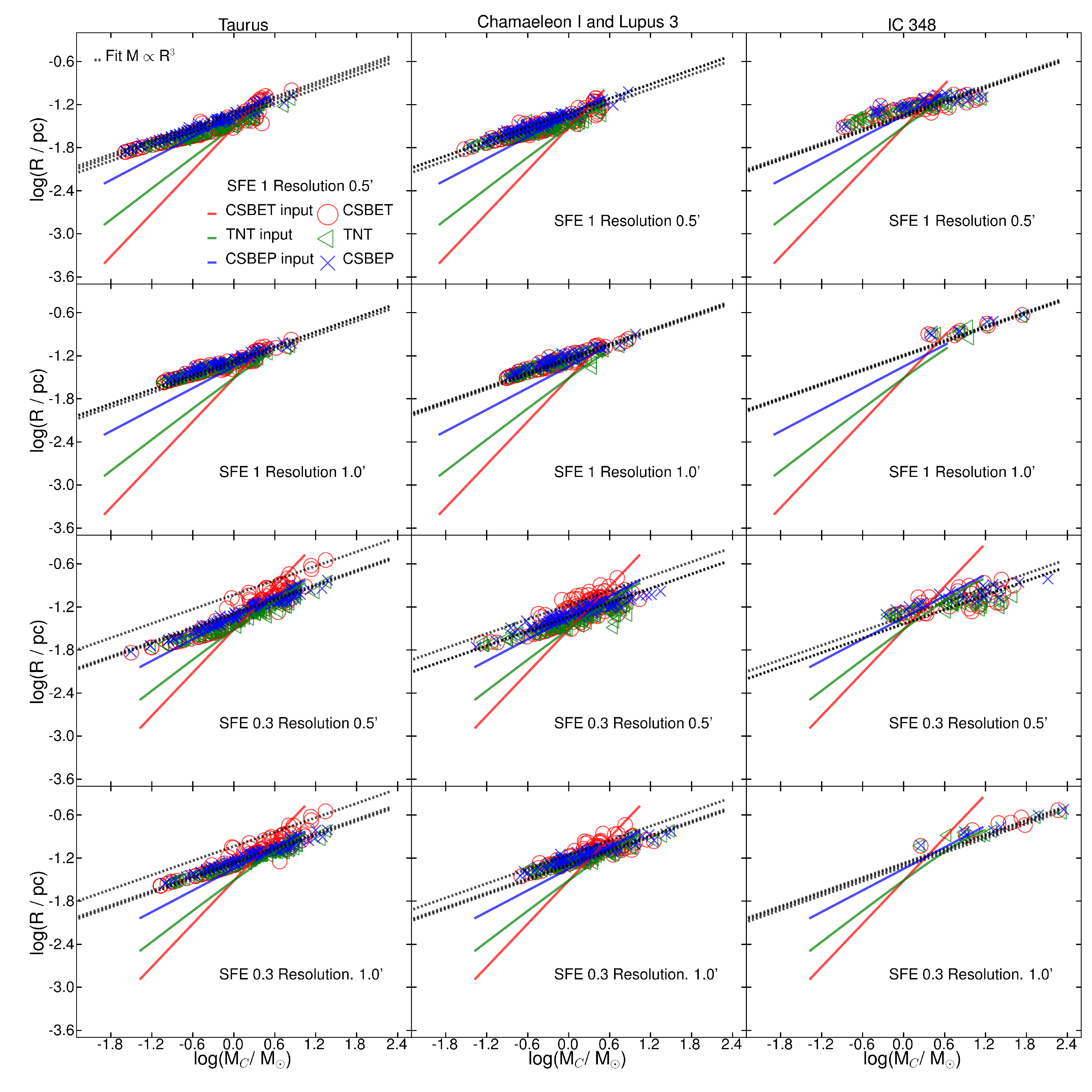} 
\caption{\small Each column shows for one region the log of the core radius against the log of the core mass M$_C$. ChaI and Lupus 3 were regrouped as their properties of crowding are much alike. Along each column, each map corresponds respectively to a different choice of parameters (SFE,resolution) and displays the results for each model. The solid lines represent the expected radii given the initial YSO mass distribution of the group, the model and SFE inputs.  The black dotted lines corresponds to the fit for each model of the power law M $\propto$ R$^3$.}
\label{fig:radius_mass}	
\end{figure}
\newpage
\begin{figure}[H]\centering
		\plotone{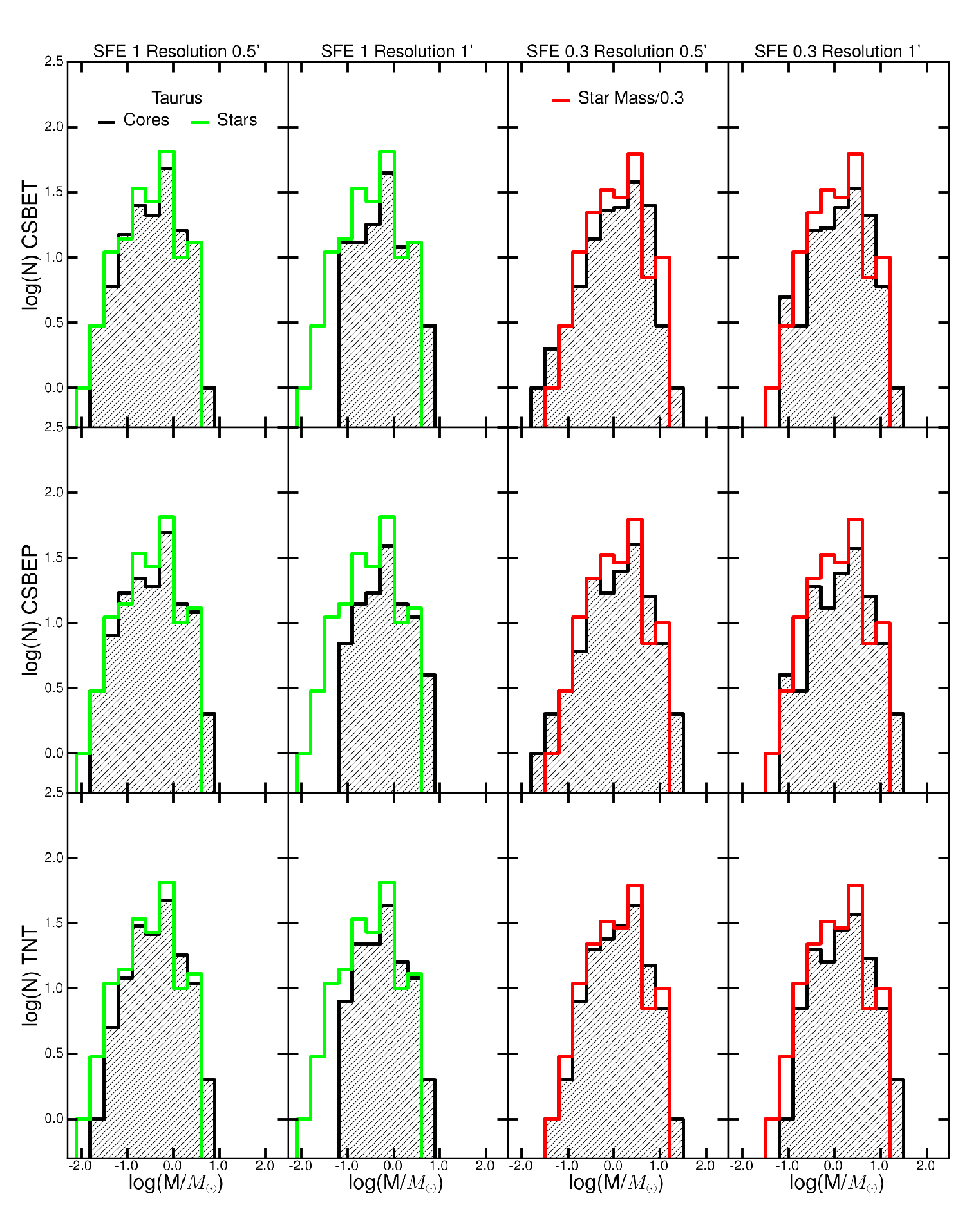}
		\caption{The distributions of masses in the Taurus region. Observed cores are shown with a black line (and shading) while the original stellar distribution is shown in green. Where the SFE is 0.3, the red line shows the stellar distribution shifted by this efficiency.
Each row corresponds respectively from top to bottom to the CSBET model, CSBEP model and TNT model. Along each row, each figure corresponds respectively from left to right to (SFE = 1.0, resolution = 0.5'),(SFE = 1.0, resolution = 1'),(SFE = 0.3, resolution = 0.5') and (SFE = 0.3, resolution = 1')}
		\label{fig:taurushisto}
\end{figure}
\newpage
\begin{figure}[H]\centering
		\plotone{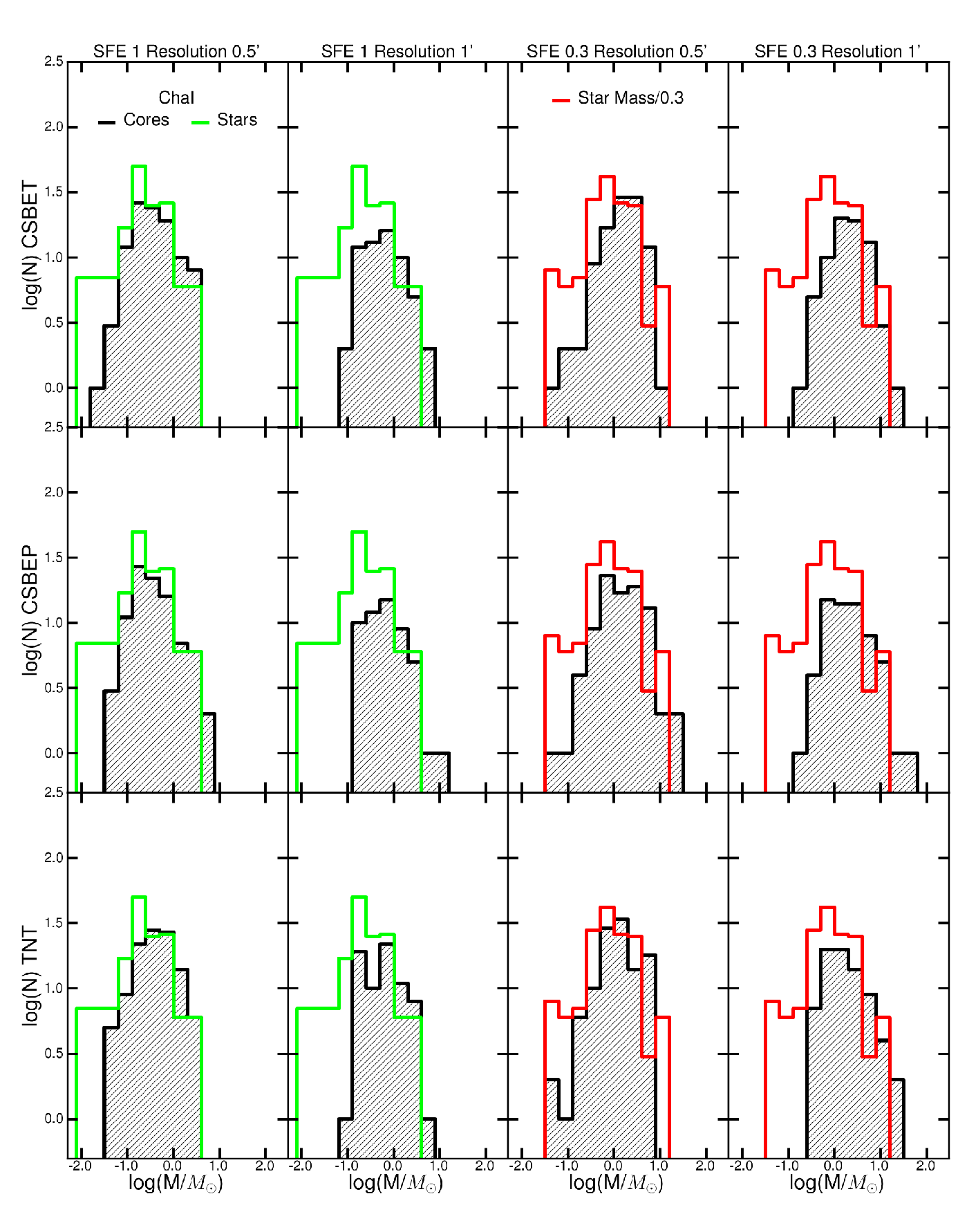}
		\caption{The distributions of masses in the Chamaeleon I region. Observed cores are shown with a black line (and shading) while the original stellar distribution is shown in green. Where the SFE is 0.3, the red line shows the stellar distribution shifted by this efficiency.
Each row corresponds respectively from top to bottom to the CSBET model, CSBEP model and TNT model. Along each row, each figure corresponds respectively from left to right to (SFE = 1.0, resolution = 0.5'),(SFE = 1.0, resolution = 1'),(SFE = 0.3, resolution = 0.5') and (SFE = 0.3, resolution = 1')}
		\label{fig:ChaIhisto}
\end{figure}
\newpage
\begin{figure}[H]\centering
		\plotone{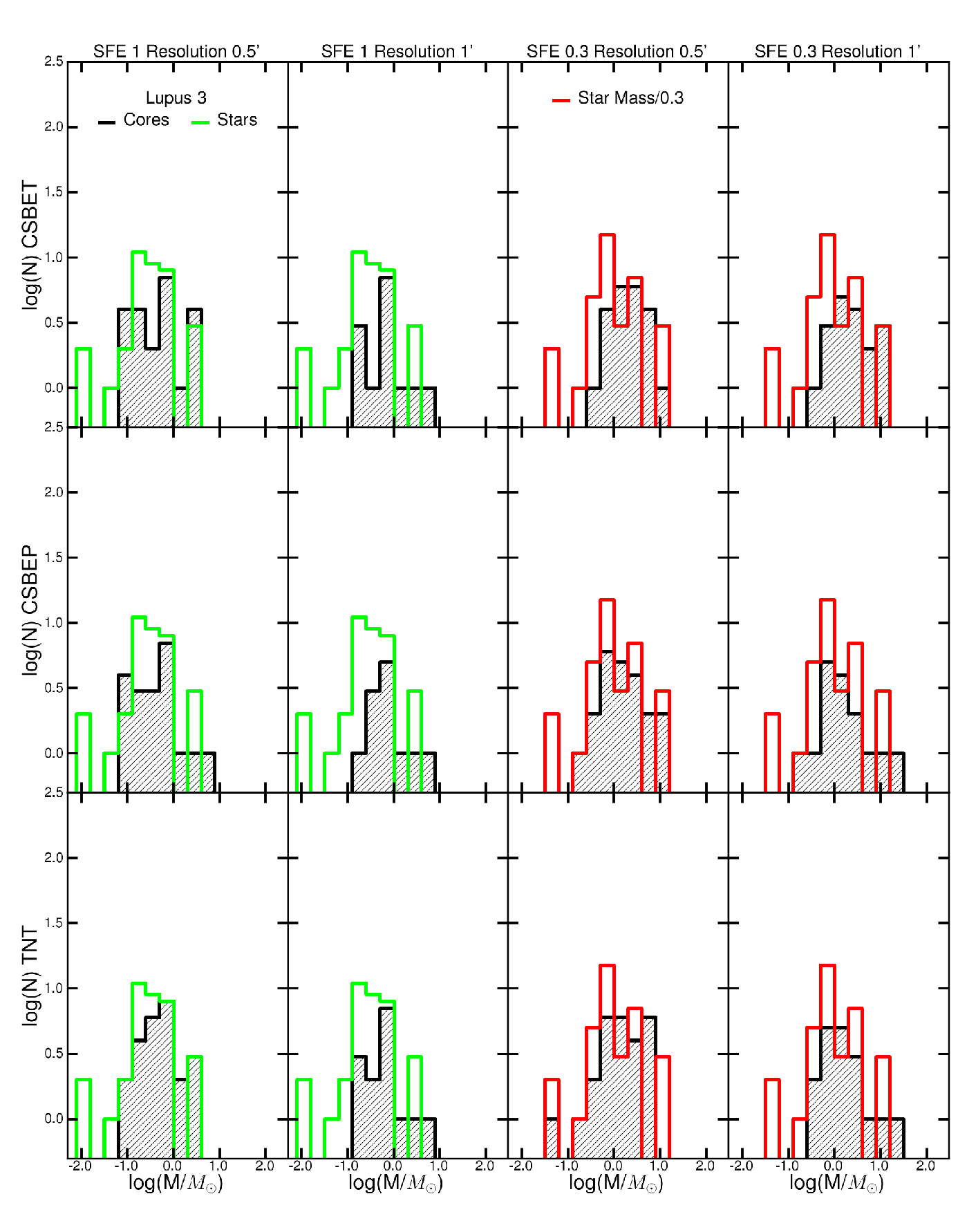}
		\caption{The distributions of masses in the Lupus 3 region. Observed cores are shown with a black line (and shading) while the original stellar distribution is shown in green. Where the SFE is 0.3, the red line shows the stellar distribution shifted by this efficiency.
Each row corresponds respectively from top to bottom to the CSBET model, CSBEP model and TNT model. Along each row, each figure corresponds respectively from left to right to (SFE = 1.0, resolution = 0.5'),(SFE = 1.0, resolution = 1'),(SFE = 0.3, resolution = 0.5') and (SFE = 0.3, resolution = 1')}
		\label{fig:lupushisto}
\end{figure}
\newpage
\begin{figure}[H]\centering
		\plotone{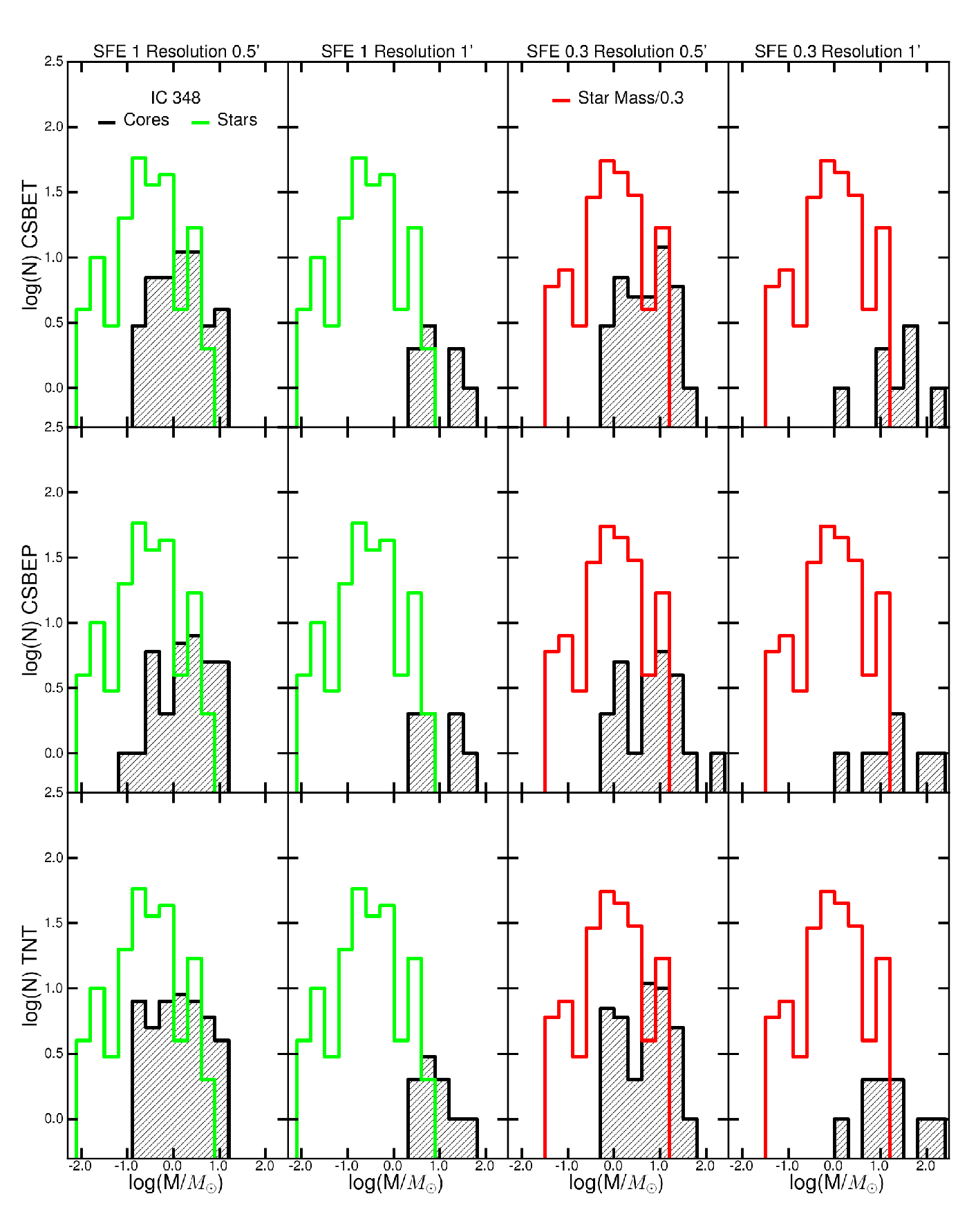}
		\caption{The distributions of masses in the IC 348 region. Observed cores are shown with a black line (and shading) while the original stellar distribution is shown in green. Where the SFE is 0.3, the red line shows the stellar distribution shifted by this efficiency.
Each row corresponds respectively from top to bottom to the CSBET model, CSBEP model and TNT model. Along each row, each figure corresponds respectively from left to right to (SFE = 1.0, resolution = 0.5'),(SFE = 1.0, resolution = 1'),(SFE = 0.3, resolution = 0.5') and (SFE = 0.3, resolution = 1')}
		\label{fig:ic348histo}
\end{figure}
\begin{figure}[H]
\centering
\epsscale{1}
		\plotone{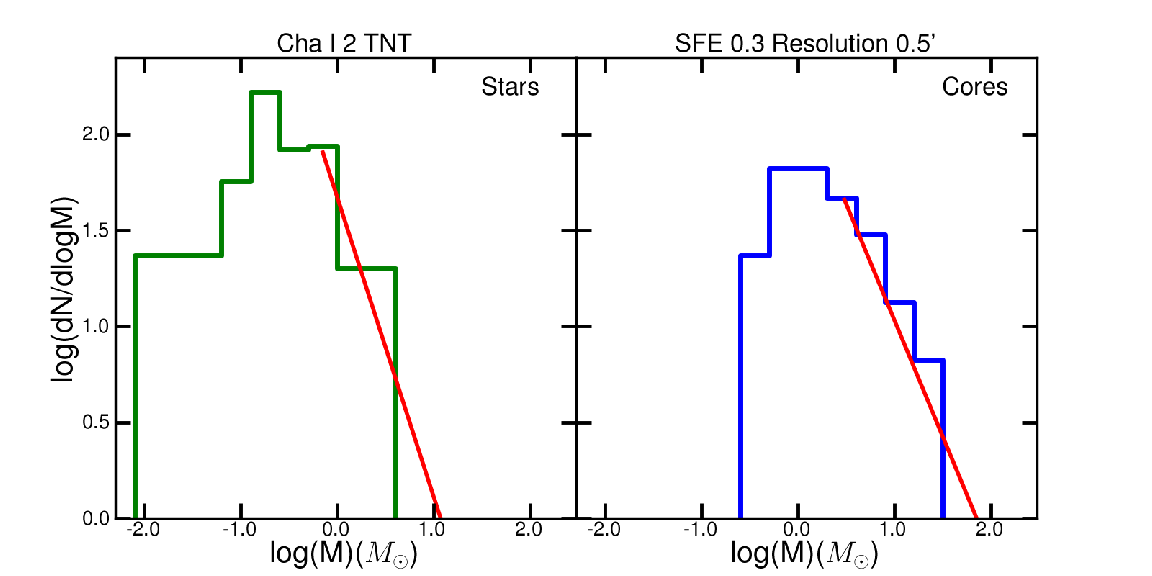}
		\caption{MLE fitting of the stellar (left) and core (right) mass function in Cha I for the TNT model, with a SFE 0.3 and a resolution 1'. The values fit for the stellar mass function are M$_{Break}$ = 0.49 M$_\odot$ and $\alpha$ = 2.56. The values fit for the observed core mass function are M$_{Break}$ = 1.46 M$_\odot$ and $\alpha$ = 2.21.}
		\label{fig:fit_quality}
\end{figure}
\begin{landscape}
\begin{figure}[H]\centering
\epsscale{0.75}
		\plotone{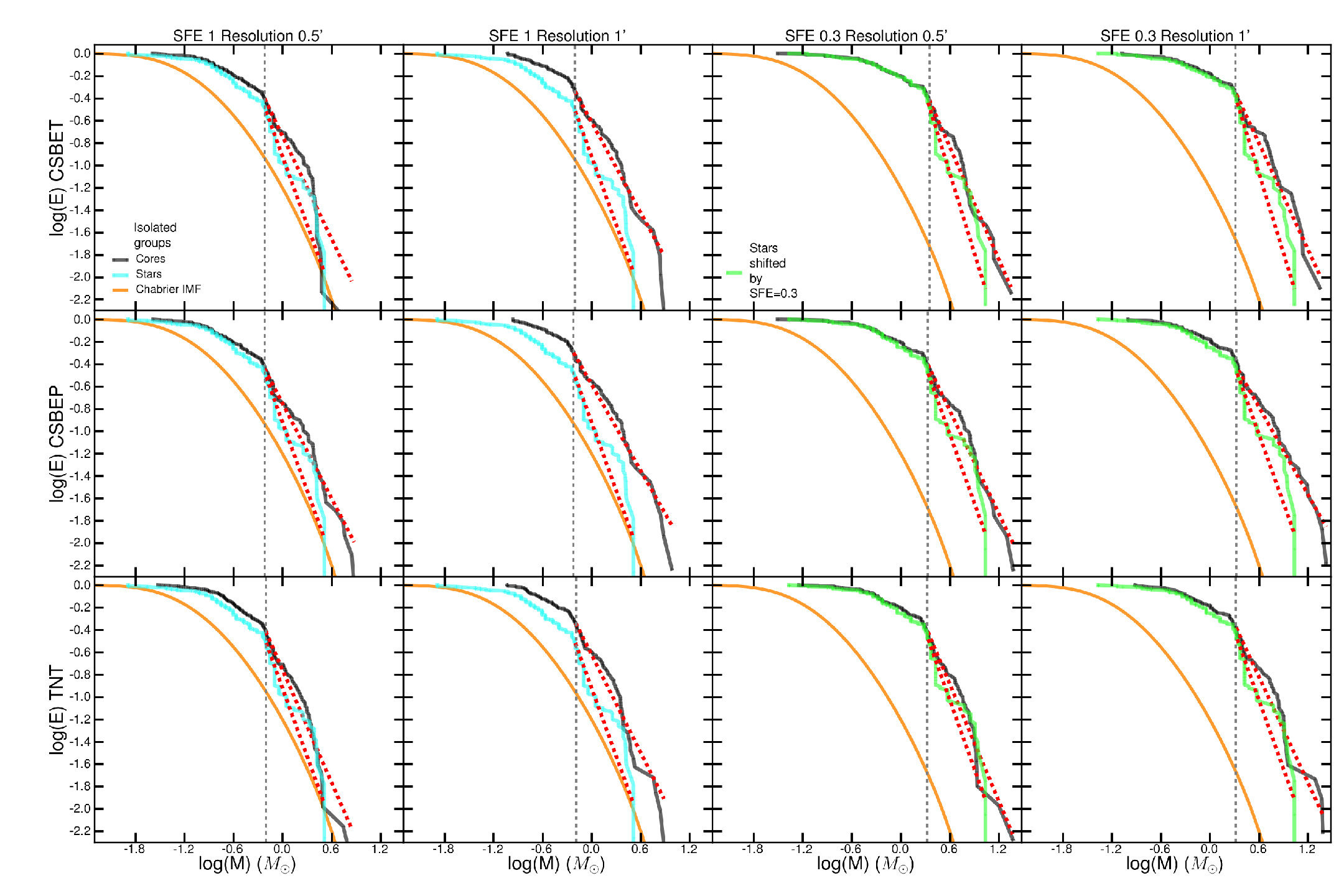}
		\caption{\footnotesize The cumulative mass distribution (fraction of stars with masses greater than M) for isolated groups. The solid back line shows the observed core mass distribution, while the solid light blue line shows the stellar mass distribution. For a SFE of 0.3, the solid green line shows the stellar mass distribution shifted by the SFE. For comparison, the solid orange line shows the Chabrier IMF \citep[see][]{Chabrier2003}.
The red dotted lines indicated the best power-law fit for the high mass tail of core and stellar distributions. Each row corresponds to the CSBET model, CSBEP model and TNT model (top to bottom). Along each row, each figure corresponds to (SFE = 1.0, resolution = 0.5'),(SFE = 1.0, resolution = 1'),(SFE = 0.3, resolution = 0.5') and (SFE = 0.3, resolution = 1') (left to right).}
		\label{fig:cumulo_I}
\end{figure}
\begin{figure}[H]\centering
\epsscale{0.75}
		\plotone{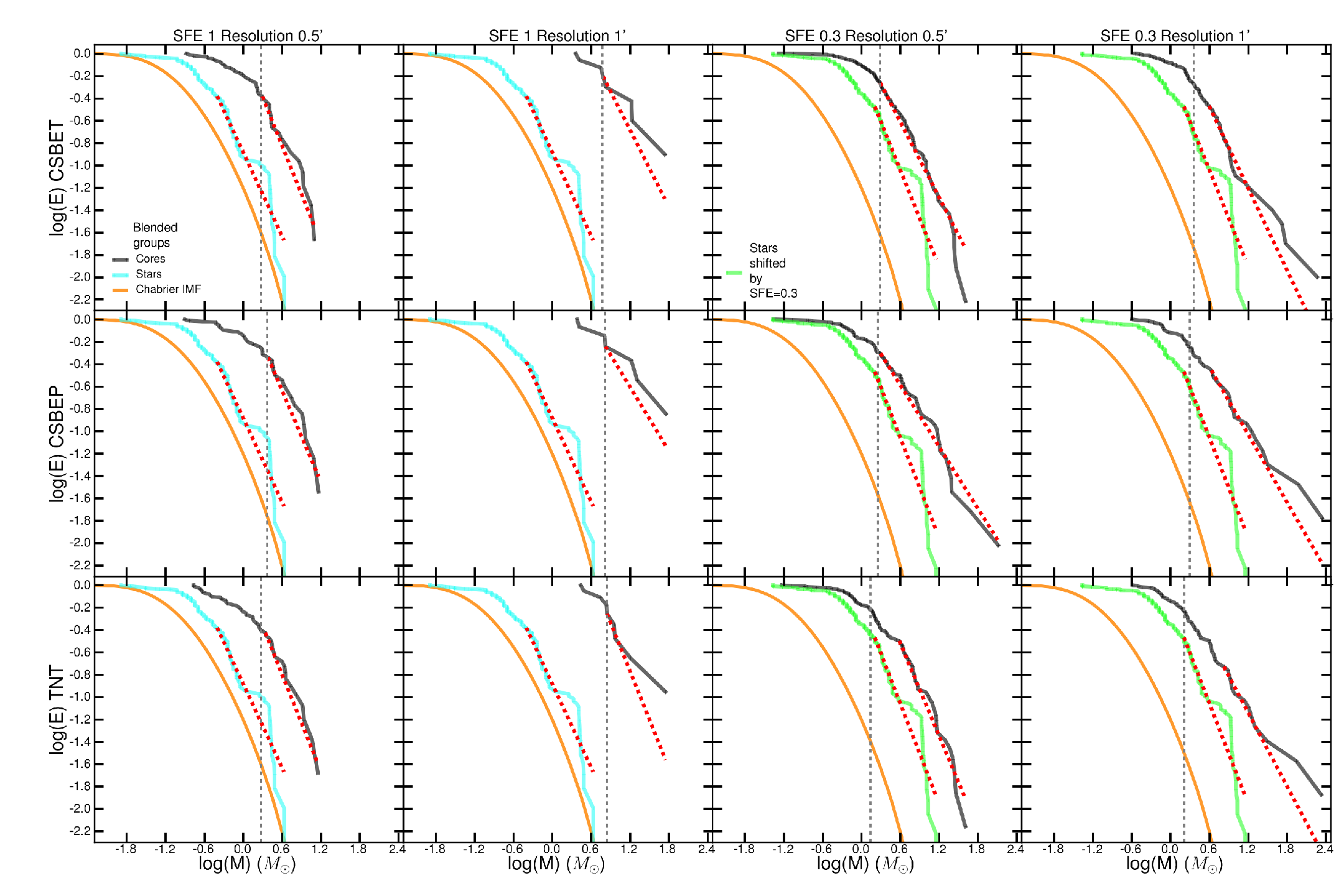}
		\caption{\footnotesize The cumulative mass distribution (fraction of stars with masses greater than M) for blended groups. The solid back line shows the observed core mass distribution, while the solid light blue line shows the stellar mass distribution. For a SFE of 0.3, the solid green line shows the stellar mass distribution shifted by the SFE. For comparison, the solid orange line shows the Chabrier IMF \citep[see][]{Chabrier2003}.
The red dotted lines indicated the best power-law fit for the high mass tail of core and stellar distributions. Each row corresponds to the CSBET model, CSBEP model and TNT model (top to bottom). Along each row, each figure corresponds to (SFE = 1.0, resolution = 0.5'),(SFE = 1.0, resolution = 1'),(SFE = 0.3, resolution = 0.5') and (SFE = 0.3, resolution = 1') (left to right).}
		\label{fig:cumulo_B}
\end{figure}
\end{landscape}
\begin{figure}[H]\centering
\epsscale{0.7}
   \plotone{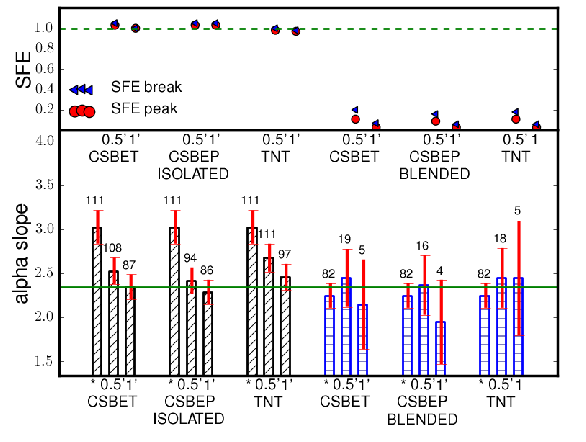}
		\plotone{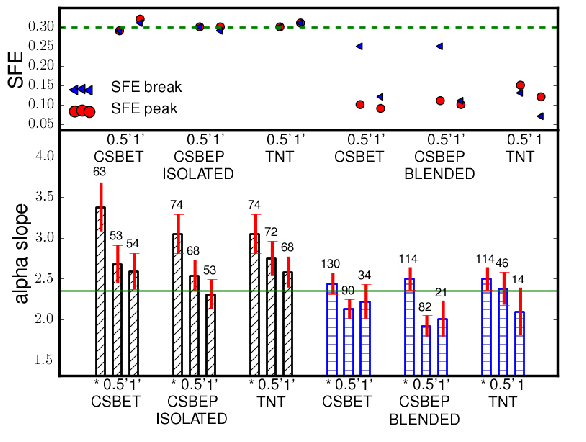}
		\caption{Star and Core mass distributions properties for an input SFE of 1.0 (\textbf{Upper graph}) and of 0.3 (\textbf{Lower graph}). For each graph are represented: \textbf{Upper panel:} The derived values of the SFE by the method of the peak mass (red circle marker) and of the break mass (blue triangle marker); the input value of the SFE is indicated by a green solid line. \textbf{Lower panel:} The derived value of the slope $\alpha$ derived by MLE fitting in the high mass range (bars), the errors on its derivation (red errorbars), the number of cores in the high mass range (Number indicated above each bar); the Salpeter value of 2.35 is indicated by a solid green line. For both panels, the values are ordered according to the degree of crowding (Isolated/ Blended, see Table 2), the input model (CSBET/ CSBEP/ TNT) and if the distribution is the initial stellar distribution (*) (for the $\alpha$ values), the distribution for a resolution of 0.5 arcmin (0.5') or the one for a resolution of 1.0 arcmin (1.0'). }
		\label{fig:sfe}
\end{figure}
\begin{figure}[H]\centering
\epsscale{0.75}
   \plotone{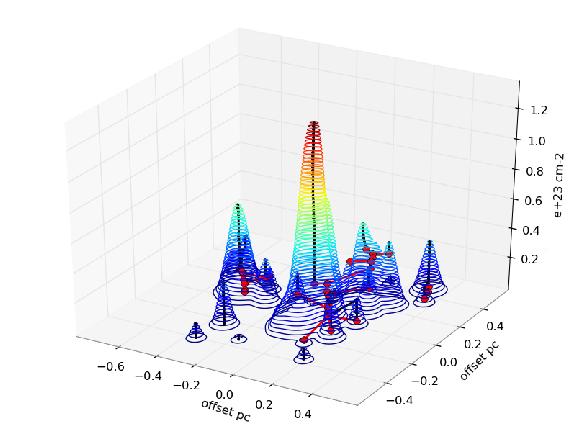}
		\plotone{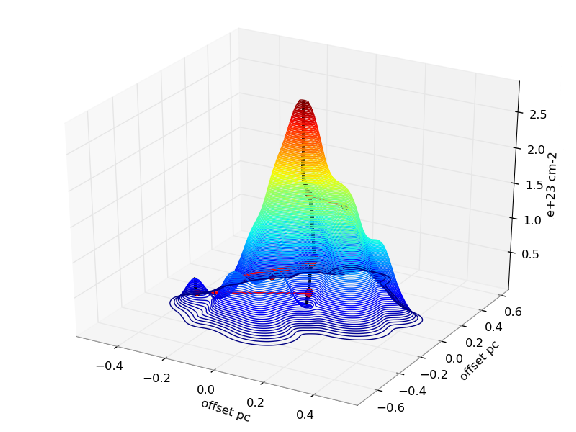}
		\caption{Contouring levels of 2$\times$(1.2$\times$10$^{21}$cm$^{-2}$) for ChaI 3 (ChaI-North) and IC 348 1 (IC 348-Main) for TNT model, SFE of 0.3 and resolution of 1.0'. The (xy) plane matches the spatial position (pc x pc unit) and the z axis matches the column density value (in 10$^{23}$cm$^{-2}$ unit). At each level, the contours of the identified clumps are drawn. The center of gravity of each current identified clump is marked as a black dot. When two previously identified clumps become contiguous at a lower level, they are represented as a global new clump and their centers of gravity are linked to the one of the new identified clump by a red line. The blending rather than the contouring levels leads to a lack of well defined peaks, necessary to the core identification.}
		\label{fig:levels}
\end{figure}
\newpage
\appendix
\section{Column density in terms of mass and projected radius\label{appendix:models}}
\par The initial condensation which produces a protostar is represented here by analytic expressions for the column density N as a function of total mass M and projected radius b, for the truncated isothermal sphere of \cite{Bonnor1956} and \cite{Ebert1955}, hereafter BE, and for the `TNT' density profile described by \cite{Myers2010}.
\par The density profile of the self-gravitating isothermal sphere is approximated to a high degree of accuracy by the expression of \cite{Natarajan1997}.
\begin{equation}
	n = n_0\left(\frac{C}{c^2 + \xi^2}-\frac{D}{d^2 + \xi^2}\right)
	\label{eq:nCSBE}
\end{equation}
where $n_0$ is the central maximum density,
\begin{equation}
\xi = \frac{r}{a}
\end{equation}
is the dimensionless radius, r is the spherical radius and a is the thermal scale length
\begin{equation}
a \equiv \frac{\sigma}{(4\pi Gmnn_0)^{1/2}}
\label{eq:a}
\end{equation}
for velocity dispersion $\sigma$. Here, G is the gravitational constant and m is the mean mass per particle, with C=50, D=48,$c^2$=10 and $d^2$=12.
\par Integration of the density in equation \ref{eq:nCSBE} yields expressions for the mass M within radius r, and for the column density N at projected radius b. The mass is
\begin{equation}
M = \frac{\sigma^3\mu}{(4\pi mn_0)^{1/2}G^{3/2}}
\label{eq:MCSBE}
\end{equation}
where the dimensionless mass is
\begin{equation}
\mu \equiv C\left(\xi - Arctan\left(\frac{\xi}{c}\right)\right) - D\left(\xi - Arctan\left(\frac{\xi}{d}\right)\right)
\label{eq:mu}
\end{equation}
If the sphere is truncated at a fixed radius R, the corresponding dimensionless radius is denoted $\xi_{max}\equiv R/a$. If the sphere is critically stable, then $\xi_{max} = 6.46$ and equation \ref{eq:mu} yields $\mu = 15.85$. This case is denoted a CSBE sphere.
\par The column density through a truncated sphere of total mass M within radius R is then given by
\begin{equation}
N = \frac{\sigma^4\mu\nu}{2\pi mG^2M}
\end{equation}
where
\begin{equation}
\nu \equiv \left[\frac{C}{\gamma}Arctan\left(\frac{(\xi_{max}^2-\beta^2)^{1/2}}{\gamma}\right)-\frac{D}{\delta}Arctan\left(\frac{(\xi_{max}^2-\beta^2)^{1/2}}{\delta}\right)\right]
\label{eq:nu}
\end{equation}
where
\begin{equation}
\beta \equiv \frac{b}{a}
\end{equation}
\begin{equation}
\gamma^2 \equiv c^2 + \beta^2
\end{equation}
and
\begin{equation}
\delta^2 \equiv d^2 + \beta^2
\end{equation}
Note that in equation \ref{eq:nu}, $0\leq\beta\leq\xi_{max}$. It is useful to write the scale length in terms of the total mass by eliminating $n_0$ from equations \ref{eq:a} and \ref{eq:MCSBE}, giving
\begin{equation}
a = \frac{GM}{\mu\sigma^2}
\end{equation}
\begin{center}
ie :   \textbf{a $\propto$ M} if $\sigma$ (ie T) is set to a fixed value (CSBET model).
\end{center}
or
\begin{equation}
a = \left(\frac{G}{4\pi P}\right)^{0.25}\left(\frac{M}{\mu}\right)^{0.5}
\end{equation} 
\begin{center}
ie :   \textbf{a $\propto$ M$^{0.5}$} if P is set to a fixed value (CSBEP model).
\end{center}
\par Another density profile is obtained from the assumption that infall is equally likely to stop at any moment and from the requirement that the resulting distribution of protostar masses follows the initial mass function (IMF). The profile ressembles a superposition of `core' and `clump' density profiles. It is also similar to the thermal and nonthermal (TNT) model of \cite{Myers1992} and to the two-component turbulent core model of \cite{McKee2003},
\begin{equation}
	n = Ar^{-2} + Br^{-2/3}
	\label{eq:nTNT}
\end{equation}

where A=34 pc$^2$cm$^{-3}$ and B=2700 pc$^{2/3}$cm$^{-3}$ (see \cite{Myers2010}).

\par The mass M within radius R is obtained by integrating equation \ref{eq:nTNT},
\begin{equation}
	M = 4\pi m\left(AR + \frac{3BR^{7/3}}{7}\right)
\end{equation}

\par Similarly, the column density through a TNT sphere of maximum radius R is
\begin{equation}
	N = \frac{2A(1 + s^2)^{1/2}}{R}Arctan(s) +2B\left(\frac{R}{(1 + s^2)^{1/2}}\right)^{1/3}\int^s_0d\zeta(1 + \zeta^2)^{-1/3}
\end{equation}

where s is a dimensionless variable, $0\leq s\leq +\infty$, and where the projected radius is
\begin{equation}
	b = R/(1 + s^2)^{1/2}
\end{equation}
\newpage

\begin{longtable}{llllccccc}
\caption[Stars and Cores properties]{Star and Core Properties} \label{tab:globaloutput} 
\\
\hline\hline\\[-2ex]
\multicolumn{1}{l}{\footnotesize Group \#}&
\multicolumn{1}{l}{\footnotesize Name}&
\multicolumn{1}{l}{\footnotesize Model}&
\multicolumn{1}{l}{\footnotesize Param.\tablenotemark{1}}&
\multicolumn{1}{c}{\footnotesize N$_{S,tot}$/N$_{C,tot}$\tablenotemark{2}}&
\multicolumn{1}{c}{\footnotesize M$_{S,tot}$/M$_{C,tot}$\tablenotemark{2}}&
\multicolumn{1}{c}{$\beta_{init}$/$\beta_{deriv}$\tablenotemark{3}}&
\multicolumn{1}{c}{\footnotesize F\tablenotemark{4}}&
\multicolumn{1}{c}{f\tablenotemark{5}}
\\
\multicolumn{1}{l}{}&
\multicolumn{1}{l}{}&
\multicolumn{1}{l}{}&
\multicolumn{1}{l}{}&
\multicolumn{1}{c}{}&
\multicolumn{1}{c}{\footnotesize(M$_\odot$/M$_\odot$)}&
\multicolumn{1}{c}{}&
\multicolumn{1}{c}{}&
\multicolumn{1}{c}{}
\\[0.5ex]
\hline
\\[-1.8ex]
\endfirsthead
\multicolumn{9}{c}{{\tablename} \thetable{} -- Continued} \\[0.5ex]
\hline\hline\\[-2ex]
\multicolumn{1}{l}{\footnotesize Group \#}&
\multicolumn{1}{l}{\footnotesize Name}&
\multicolumn{1}{l}{\footnotesize Model}&
\multicolumn{1}{l}{\footnotesize Param.\tablenotemark{1}}&
\multicolumn{1}{c}{\footnotesize N$_{S,tot}$/N$_{C,tot}$\tablenotemark{2}}&
\multicolumn{1}{c}{\footnotesize M$_{S,tot}$/M$_{C,tot}$\tablenotemark{2}}&
\multicolumn{1}{c}{$\beta_{init}$/$\beta_{deriv}$\tablenotemark{3}}&
\multicolumn{1}{c}{\footnotesize F\tablenotemark{4}}&
\multicolumn{1}{c}{f\tablenotemark{5}}
\\
\multicolumn{1}{l}{}&
\multicolumn{1}{l}{}&
\multicolumn{1}{l}{}&
\multicolumn{1}{l}{}&
\multicolumn{1}{c}{}&
\multicolumn{1}{c}{\footnotesize(M$_\odot$/M$_\odot$)}&
\multicolumn{1}{c}{}&
\multicolumn{1}{c}{}&
\multicolumn{1}{c}{}
\\[0.5ex]
\hline
\\[-1.8ex]
\endhead
\endlastfoot
\textbf{\scriptsize Taurus1}&\textbf{\scriptsize B209}&\textbf{\scriptsize CSBET}&\footnotesize(1,0.5')&\footnotesize20/15&\footnotesize10.64/10.38&\footnotesize1.0/3.1&\footnotesize1.33&\footnotesize2.87\\
--&--&--&\footnotesize(1,1')&\footnotesize--/14&\footnotesize--/9.98&\footnotesize1.0/3.8&\footnotesize1.43&\footnotesize2.25\\
--&--&--&\footnotesize(0.3,0.5')&\footnotesize--/15&\footnotesize--/34.80&\footnotesize1.0/2.4&\footnotesize1.33&\footnotesize1.57\\
--&--&--&\footnotesize(0.3,1')&\footnotesize--/13&\footnotesize--/34.33&\footnotesize1.0/2.7&\footnotesize1.54&\footnotesize1.42\\
\textbf{\scriptsize Taurus1}&\textbf{\scriptsize B209}&\textbf{\scriptsize CSBEP}&\footnotesize(1,0.5')&\footnotesize20/15&\footnotesize10.64/10.27&\footnotesize2.0/3.8&\footnotesize1.33&\footnotesize2.40\\
--&--&--&\footnotesize(1,1')&\footnotesize--/11&\footnotesize--/9.76&\footnotesize2.0/4.6&\footnotesize1.82&\footnotesize2.36\\
--&--&--&\footnotesize(0.3,0.5')&\footnotesize--/14&\footnotesize--/34.90&\footnotesize2.0/2.9&\footnotesize1.43&\footnotesize1.76\\
--&--&--&\footnotesize(0.3,1')&\footnotesize--/12&\footnotesize--/34.36&\footnotesize2.0/3.7&\footnotesize1.67&\footnotesize1.56\\
\textbf{\scriptsize Taurus1}&\textbf{\scriptsize B209}&\textbf{\scriptsize TNT}&\footnotesize(1,0.5')&\footnotesize20/17&\footnotesize10.64/10.77&\footnotesize1.4/3.3&\footnotesize1.18&\footnotesize2.80\\
--&--&--&\footnotesize(1,1')&\footnotesize--/15&\footnotesize--/10.34&\footnotesize1.4/3.3&\footnotesize1.33&\footnotesize2.05\\
--&--&--&\footnotesize(0.3,0.5')&\footnotesize--/16&\footnotesize--/34.13&\footnotesize1.6/2.9&\footnotesize1.25&\footnotesize1.84\\
--&--&--&\footnotesize(0.3,1')&\footnotesize--/14&\footnotesize--/33.59&\footnotesize1.6/3.4&\footnotesize1.43&\footnotesize1.50\\
\textbf{\scriptsize Taurus2}&\textbf{\scriptsize L1495E}&\textbf{\scriptsize CSBET}&\footnotesize(1,0.5')&\footnotesize30/27&\footnotesize15.53/15.10&\footnotesize1.0/3.1&\footnotesize1.11&\footnotesize2.36\\
--&--&--&\footnotesize(1,1')&\footnotesize--/21&\footnotesize--/14.21&\footnotesize1.0/3.5&\footnotesize1.43&\footnotesize2.00\\
--&--&--&\footnotesize(0.3,0.5')&\footnotesize--/24&\footnotesize--/49.66&\footnotesize1.0/2.3&\footnotesize1.25&\footnotesize1.44\\
--&--&--&\footnotesize(0.3,1')&\footnotesize--/22&\footnotesize--/49.12&\footnotesize1.0/2.7&\footnotesize1.36&\footnotesize1.23\\
\textbf{\scriptsize Taurus2}&\textbf{\scriptsize L1495E}&\textbf{\scriptsize CSBEP}&\footnotesize(1,0.5')&\footnotesize30/27&\footnotesize15.53/14.96&\footnotesize2.0/3.4&\footnotesize1.11&\footnotesize2.06\\
--&--&--&\footnotesize(1,1')&\footnotesize--/17&\footnotesize--/13.87&\footnotesize2.0/3.7&\footnotesize1.76&\footnotesize2.10\\
--&--&--&\footnotesize(0.3,0.5')&\footnotesize--/25&\footnotesize--/50.92&\footnotesize2.0/2.7&\footnotesize1.20&\footnotesize1.58\\
--&--&--&\footnotesize(0.3,1')&\footnotesize--/23&\footnotesize--/50.11&\footnotesize2.0/3.6&\footnotesize1.30&\footnotesize1.33\\
\textbf{\scriptsize Taurus2}&\textbf{\scriptsize L1495E}&\textbf{\scriptsize TNT}&\footnotesize(1,0.5')&\footnotesize30/28&\footnotesize15.53/15.82&\footnotesize1.4/3.1&\footnotesize1.07&\footnotesize2.28\\
--&--&--&\footnotesize(1,1')&\footnotesize--/22&\footnotesize--/14.99&\footnotesize1.4/3.3&\footnotesize1.36&\footnotesize2.08\\
--&--&--&\footnotesize(0.3,0.5')&\footnotesize--/28&\footnotesize--/49.79&\footnotesize1.6/2.9&\footnotesize1.07&\footnotesize1.63\\
--&--&--&\footnotesize(0.3,1')&\footnotesize--/24&\footnotesize--/49.00&\footnotesize1.6/3.5&\footnotesize1.25&\footnotesize1.39\\
\textbf{\scriptsize Taurus3}&\textbf{\scriptsize B213}&\textbf{\scriptsize CSBET}&\footnotesize(1,0.5')&\footnotesize19/16&\footnotesize8.08/7.84&\footnotesize1.0/3.9&\footnotesize1.19&\footnotesize3.26\\
--&--&--&\footnotesize(1,1')&\footnotesize--/13&\footnotesize--/7.32&\footnotesize1.0/3.9&\footnotesize1.46&\footnotesize2.76\\
--&--&--&\footnotesize(0.3,0.5')&\footnotesize--/17&\footnotesize--/26.41&\footnotesize1.0/2.4&\footnotesize1.12&\footnotesize1.83\\
--&--&--&\footnotesize(0.3,1')&\footnotesize--/14&\footnotesize--/25.85&\footnotesize1.0/2.9&\footnotesize1.36&\footnotesize.69\\
\textbf{\scriptsize Taurus3}&\textbf{\scriptsize B213}&\textbf{\scriptsize CSBEP}&\footnotesize(1,0.5')&\footnotesize19/15&\footnotesize8.08/7.72&\footnotesize2.0/3.8&\footnotesize1.27&\footnotesize2.92\\
--&--&--&\footnotesize(1,1')&\footnotesize--/12&\footnotesize--/7.10&\footnotesize2.0/3.9&\footnotesize1.58&\footnotesize2.69\\
--&--&--&\footnotesize(0.3,0.5')&\footnotesize--/16&\footnotesize--/26.39&\footnotesize2.0/2.7&\footnotesize1.19&\footnotesize1.91\\
--&--&--&\footnotesize(0.3,1')&\footnotesize--/14&\footnotesize--/25.81&\footnotesize2.0/3.3&\footnotesize1.36&\footnotesize1.69\\
\textbf{\scriptsize Taurus3}&\textbf{\scriptsize B213}&\textbf{\scriptsize TNT}&\footnotesize(1,0.5')&\footnotesize19/17&\footnotesize8.08/8.43&\footnotesize1.4/3.2&\footnotesize1.12&\footnotesize3.23\\
--&--&--&\footnotesize(1,1')&\footnotesize--/14&\footnotesize--/7.96&\footnotesize1.4/3.6&\footnotesize1.36&\footnotesize2.58\\
--&--&--&\footnotesize(0.3,0.5')&\footnotesize--/17&\footnotesize--/25.76&\footnotesize1.5/2.8&\footnotesize1.12&\footnotesize2.04\\
--&--&--&\footnotesize(0.3,1')&\footnotesize--/14&\footnotesize--/25.19&\footnotesize1.5/3.4&\footnotesize1.36&\footnotesize1.84\\
\textbf{\scriptsize Taurus4}&\textbf{\scriptsize L1551}&\textbf{\scriptsize CSBET}&\footnotesize(1,0.5')&\footnotesize24/18&\footnotesize22.69/22.29&\footnotesize1.0/2.8&\footnotesize1.33&\footnotesize2.47\\
--&--&--&\footnotesize(1,1')&\footnotesize--/14&\footnotesize--/21.66&\footnotesize1.0/3.2&\footnotesize1.71&\footnotesize2.10\\
--&--&--&\footnotesize(0.3,0.5')&\footnotesize--/15&\footnotesize--/73.28&\footnotesize1.0/1.9&\footnotesize1.60&\footnotesize1.45\\
--&--&--&\footnotesize(0.3,1')&\footnotesize--/14&\footnotesize--/72.62&\footnotesize1.0/2.4&\footnotesize1.71&\footnotesize1.45\\
\textbf{\scriptsize Taurus4}&\textbf{\scriptsize L1551}&\textbf{\scriptsize CSBEP}&\footnotesize(1,0.5')&\footnotesize24/18&\footnotesize22.69/22.20&\footnotesize2.0/3.0&\footnotesize1.33&\footnotesize2.23\\
--&--&--&\footnotesize(1,1')&\footnotesize--/13&\footnotesize--/21.55&\footnotesize2.0/3.8&\footnotesize1.85&\footnotesize2.19\\
--&--&--&\footnotesize(0.3,0.5')&\footnotesize--/17&\footnotesize--/74.79&\footnotesize2.0/2.5&\footnotesize1.41&\footnotesize1.45\\
--&--&--&\footnotesize(0.3,1')&\footnotesize--/14&\footnotesize--/74.06&\footnotesize2.0/3.2&\footnotesize1.71&\footnotesize1.74\\
\textbf{\scriptsize Taurus4}&\textbf{\scriptsize L1551}&\textbf{\scriptsize TNT}&\footnotesize(1,0.5')&\footnotesize24/18&\footnotesize22.69/22.30&\footnotesize1.4/3.1&\footnotesize1.33&\footnotesize2.59\\
--&--&--&\footnotesize(1,1')&\footnotesize--/15&\footnotesize--/21.72&\footnotesize1.4/3.6&\footnotesize1.60&\footnotesize2.32\\
--&--&--&\footnotesize(0.3,0.5')&\footnotesize--/18&\footnotesize--/73.32&\footnotesize2.0/2.7&\footnotesize1.33&\footnotesize1.53\\
--&--&--&\footnotesize(0.3,1')&\footnotesize--/16&\footnotesize--/72.62&\footnotesize2.0/3.2&\footnotesize1.50&\footnotesize1.56\\
\textbf{\scriptsize Taurus5}&\textbf{\scriptsize L1529}&\textbf{\scriptsize CSBET}&\footnotesize(1,0.5')&\footnotesize14/9&\footnotesize8.25/8.04&\footnotesize1.0/3.9&\footnotesize1.56&\footnotesize4.27\\
--&--&--&\footnotesize(1,1')&\footnotesize--/7&\footnotesize--/7.69&\footnotesize1.0/4.6&\footnotesize2.00&\footnotesize3.68\\
--&--&--&\footnotesize(0.3,0.5')&\footnotesize--/10&\footnotesize--/26.93&\footnotesize1.0/2.3&\footnotesize1.40&\footnotesize2.25\\
--&--&--&\footnotesize(0.3,1')&\footnotesize--/10&\footnotesize--/26.44&\footnotesize1.0/3.0&\footnotesize1.40&\footnotesize2.06\\
\textbf{\scriptsize Taurus5}&\textbf{\scriptsize L1529}&\textbf{\scriptsize CSBEP}&\footnotesize(1,0.5')&\footnotesize14/10&\footnotesize8.25/7.99&\footnotesize2.0/3.9&\footnotesize1.40&\footnotesize4.02\\
--&--&--&\footnotesize(1,1')&\footnotesize--/7&\footnotesize--/7.59&\footnotesize2.0/4.3&\footnotesize2.00&\footnotesize3.34\\
--&--&--&\footnotesize(0.3,0.5')&\footnotesize--/10&\footnotesize--/27.03&\footnotesize2.0/2.7&\footnotesize1.40&\footnotesize2.41\\
--&--&--&\footnotesize(0.3,1')&\footnotesize--/9&\footnotesize--/26.50&\footnotesize2.0/3.5&\footnotesize1.56&\footnotesize1.88\\
\textbf{\scriptsize Taurus5}&\textbf{\scriptsize L1529}&\textbf{\scriptsize TNT}&\footnotesize(1,0.5')&\footnotesize14/10&\footnotesize8.25/8.50&\footnotesize1.4/3.3&\footnotesize1.40&\footnotesize4.73\\
--&--&--&\footnotesize(1,1')&\footnotesize--/8&\footnotesize--/8.14&\footnotesize1.4/4.5&\footnotesize1.75&\footnotesize3.38\\
--&--&--&\footnotesize(0.3,0.5')&\footnotesize--/10&\footnotesize--/26.44&\footnotesize1.6/2.8&\footnotesize1.40&\footnotesize2.74\\
--&--&--&\footnotesize(0.3,1')&\footnotesize--/10&\footnotesize--/26.03&\footnotesize1.6/3.4&\footnotesize1.40&\footnotesize2.34\\
\textbf{\scriptsize Taurus6}&\textbf{\scriptsize L1536}&\textbf{\scriptsize CSBET}&\footnotesize(1,0.5')&\footnotesize31/27&\footnotesize17.67/17.25&\footnotesize1.0/3.6&\footnotesize1.15&\footnotesize2.86\\
--&--&--&\footnotesize(1,1')&\footnotesize--/16&\footnotesize--/16.31&\footnotesize1.0/3.6&\footnotesize1.94&\footnotesize2.48\\
--&--&--&\footnotesize(0.3,0.5')&\footnotesize--/26&\footnotesize--/57.15&\footnotesize1.0/2.1&\footnotesize1.19&\footnotesize1.42\\
--&--&--&\footnotesize(0.3,1')&\footnotesize--/23&\footnotesize--/56.39&\footnotesize1.0/2.8&\footnotesize1.35&\footnotesize1.40\\
\textbf{\scriptsize Taurus6}&\textbf{\scriptsize L1536}&\textbf{\scriptsize CSBEP}&\footnotesize(1,0.5')&\footnotesize31/25&\footnotesize17.67/17.10&\footnotesize2.0/3.4&\footnotesize1.24&\footnotesize2.64\\
--&--&--&\footnotesize(1,1')&\footnotesize--/16&\footnotesize--/16.16&\footnotesize2.0/3.8&\footnotesize1.94&\footnotesize2.33\\
--&--&--&\footnotesize(0.3,0.5')&\footnotesize--/24&\footnotesize--/57.98&\footnotesize2.0/2.7&\footnotesize1.29&\footnotesize1.69\\
--&--&--&\footnotesize(0.3,1')&\footnotesize--/20&\footnotesize--/57.01&\footnotesize2.0/3.5&\footnotesize1.55&\footnotesize1.59\\
\textbf{\scriptsize Taurus6}&\textbf{\scriptsize L1536}&\textbf{\scriptsize TNT}&\footnotesize(1,0.5')&\footnotesize31/25&\footnotesize17.67/17.77&\footnotesize1.4/3.2&\footnotesize1.24&\footnotesize3.04\\
--&--&--&\footnotesize(1,1')&\footnotesize--/20&\footnotesize--/17.03&\footnotesize1.4/3.7&\footnotesize1.55&\footnotesize2.30\\
--&--&--&\footnotesize(0.3,0.5')&\footnotesize--/26&\footnotesize--/57.11&\footnotesize1.7/2.9&\footnotesize1.19&\footnotesize1.81\\
--&--&--&\footnotesize(0.3,1')&\footnotesize--/23&\footnotesize--/56.29&\footnotesize1.7/3.3&\footnotesize1.35&\footnotesize1.66\\
\textbf{\scriptsize Taurus7}&\textbf{\scriptsize L1527}&\textbf{\scriptsize CSBET}&\footnotesize(1,0.5')&\footnotesize24/21&\footnotesize16.13/15.69&\footnotesize1.0/3.0&\footnotesize1.14&\footnotesize2.98\\
--&--&--&\footnotesize(1,1')&\footnotesize--/18&\footnotesize--/14.98&\footnotesize1.0/3.3&\footnotesize1.33&\footnotesize2.12\\
--&--&--&\footnotesize(0.3,0.5')&\footnotesize--/20&\footnotesize--/51.49&\footnotesize1.0/1.9&\footnotesize1.20&\footnotesize1.39\\
--&--&--&\footnotesize(0.3,1')&\footnotesize--/19&\footnotesize--/50.73&\footnotesize1.0/2.4&\footnotesize1.26&\footnotesize1.29\\
\textbf{\scriptsize Taurus7}&\textbf{\scriptsize L1527}&\textbf{\scriptsize CSBEP}&\footnotesize(1,0.5')&\footnotesize24/21&\footnotesize16.13/15.55&\footnotesize2.0/3.0&\footnotesize1.14&\footnotesize2.61\\
--&--&--&\footnotesize(1,1')&\footnotesize--/17&\footnotesize--/14.74&\footnotesize2.0/3.6&\footnotesize1.41&\footnotesize1.92\\
--&--&--&\footnotesize(0.3,0.5')&\footnotesize--/21&\footnotesize--/52.77&\footnotesize2.0/2.3&\footnotesize1.14&\footnotesize1.56\\
--&--&--&\footnotesize(0.3,1')&\footnotesize--/19&\footnotesize--/51.83&\footnotesize2.0/3.0&\footnotesize1.26&\footnotesize1.31\\
\textbf{\scriptsize Taurus7}&\textbf{\scriptsize L1527}&\textbf{\scriptsize TNT}&\footnotesize(1,0.5')&\footnotesize24/22&\footnotesize16.13/15.97&\footnotesize1.4/2.9&\footnotesize1.09&\footnotesize2.92\\
--&--&--&\footnotesize(1,1')&\footnotesize--/18&\footnotesize--/15.24&\footnotesize1.4/3.4&\footnotesize1.33&\footnotesize2.14\\
--&--&--&\footnotesize(0.3,0.5')&\footnotesize--/21&\footnotesize--/51.96&\footnotesize1.7/2.5&\footnotesize1.14&\footnotesize1.78\\
--&--&--&\footnotesize(0.3,1')&\footnotesize--/20&\footnotesize--/51.14&\footnotesize1.7/3.0&\footnotesize1.20&\footnotesize1.56\\
\textbf{\scriptsize Taurus8}&\textbf{\scriptsize L1517}&\textbf{\scriptsize CSBET}&\footnotesize(1,0.5')&\footnotesize16/15&\footnotesize12.27/11.94&\footnotesize1.0/2.4&\footnotesize1.07&\footnotesize2.31\\
--&--&--&\footnotesize(1,1')&\footnotesize--/13&\footnotesize--/11.48&\footnotesize1.0/3.1&\footnotesize1.23&\footnotesize1.90\\
--&--&--&\footnotesize(0.3,0.5')&\footnotesize--/13&\footnotesize--/38.58&\footnotesize1.0/1.7&\footnotesize1.23&\footnotesize0.97\\
--&--&--&\footnotesize(0.3,1')&\footnotesize--/12&\footnotesize--/38.44&\footnotesize1.0/2.2&\footnotesize1.33&\footnotesize1.01\\
\textbf{\scriptsize Taurus8}&\textbf{\scriptsize L1517}&\textbf{\scriptsize CSBEP}&\footnotesize(1,0.5')&\footnotesize16/15&\footnotesize12.27/11.90&\footnotesize2.0/2.5&\footnotesize1.07&\footnotesize2.22\\
--&--&--&\footnotesize(1,1')&\footnotesize--/13&\footnotesize--/11.43&\footnotesize2.0/3.5&\footnotesize1.23&\footnotesize1.90\\
--&--&--&\footnotesize(0.3,0.5')&\footnotesize--/14&\footnotesize--/40.32&\footnotesize2.0/2.2&\footnotesize1.14&\footnotesize1.41\\
--&--&--&\footnotesize(0.3,1')&\footnotesize--/14&\footnotesize--/39.84&\footnotesize2.0/3.0&\footnotesize1.14&\footnotesize1.29\\
\textbf{\scriptsize Taurus8}&\textbf{\scriptsize L1517}&\textbf{\scriptsize TNT}&\footnotesize(1,0.5')&\footnotesize16/15&\footnotesize12.27/12.08&\footnotesize1.4/2.7&\footnotesize1.07&\footnotesize2.53\\
--&--&--&\footnotesize(1,1')&\footnotesize--/13&\footnotesize--/11.59&\footnotesize1.4/3.4&\footnotesize1.23&\footnotesize1.98\\
--&--&--&\footnotesize(0.3,0.5')&\footnotesize--/15&\footnotesize--/40.20&\footnotesize2.0/2.5&\footnotesize1.07&\footnotesize1.49\\
--&--&--&\footnotesize(0.3,1')&\footnotesize--/14&\footnotesize--/39.74&\footnotesize2.0/3.2&\footnotesize1.14&\footnotesize1.37\\
\textbf{\scriptsize ChaI1}&\textbf{\scriptsize ChaI}&\textbf{\scriptsize CSBET}&\footnotesize(1,0.5')&\footnotesize12/10&\footnotesize3.66/3.47&\footnotesize1.0/3.8&\footnotesize1.20&\footnotesize2.70\\
--&\scriptsize \textbf{-SW}&--&\footnotesize(1,1')&\footnotesize--/8&\footnotesize--/2.98&\footnotesize1.0/4.0&\footnotesize1.50&\footnotesize1.71\\
--&--&--&\footnotesize(0.3,0.5')&\footnotesize--/11&\footnotesize--/11.82&\footnotesize1.0/2.5&\footnotesize1.09&\footnotesize1.43\\
--&--&--&\footnotesize(0.3,1')&\footnotesize--/10&\footnotesize--/11.44&\footnotesize1.0/2.8&\footnotesize1.20&\footnotesize1.25\\
\textbf{\scriptsize ChaI1}&\textbf{\scriptsize ChaI}&\textbf{\scriptsize CSBEP}&\footnotesize(1,0.5')&\footnotesize12/10&\footnotesize3.66/3.39&\footnotesize2.0/3.9&\footnotesize1.20&\footnotesize2.31\\
--&\scriptsize \textbf{-SW}&--&\footnotesize(1,1')&\footnotesize--/6&\footnotesize--/2.77&\footnotesize2.0/4.2&\footnotesize2.00&\footnotesize1.81\\
--&--&--&\footnotesize(0.3,0.5')&\footnotesize--/10&\footnotesize--/11.78&\footnotesize2.0/2.7&\footnotesize1.20&\footnotesize1.44\\
--&--&--&\footnotesize(0.3,1')&\footnotesize--/10&\footnotesize--/11.40&\footnotesize2.0/3.4&\footnotesize1.20&\footnotesize1.22\\
\textbf{\scriptsize ChaI1}&\textbf{\scriptsize ChaI}&\textbf{\scriptsize TNT}&\footnotesize(1,0.5')&\footnotesize12/11&\footnotesize3.66/4.11&\footnotesize1.4/3.0&\footnotesize1.09&\footnotesize2.32\\
--&\scriptsize \textbf{-SW}&--&\footnotesize(1,1')&\footnotesize--/9&\footnotesize--/3.63&\footnotesize1.4/3.5&\footnotesize1.33&\footnotesize1.76\\
--&--&--&\footnotesize(0.3,0.5')&\footnotesize--/11&\footnotesize--/11.93&\footnotesize1.5/2.4&\footnotesize1.09&\footnotesize1.54\\
--&--&--&\footnotesize(0.3,1')&\footnotesize--/10&\footnotesize--/11.57&\footnotesize1.5/3.2&\footnotesize1.20&\footnotesize1.30\\
\textbf{\scriptsize ChaI2}&\textbf{\scriptsize ChaI}&\textbf{\scriptsize CSBET}&\footnotesize(1,0.5')&\footnotesize96/63&\footnotesize40.45/39.18&\footnotesize1.0/3.2&\footnotesize1.52&\footnotesize1.30\\
--&\scriptsize \textbf{-South}&--&\footnotesize(1,1')&\footnotesize--/34&\footnotesize--/37.08&\footnotesize1.0/3.3&\footnotesize2.82&\footnotesize1.22\\
--&--&--&\footnotesize(0.3,0.5')&\footnotesize--/64&\footnotesize--/131.15&\footnotesize1.0/2.8&\footnotesize1.50&\footnotesize0.66\\
--&--&--&\footnotesize(0.3,1')&\footnotesize--/41&\footnotesize--/130.86&\footnotesize1.0/3.2&\footnotesize2.34&\footnotesize0.68\\
\textbf{\scriptsize ChaI2}&\textbf{\scriptsize ChaI}&\textbf{\scriptsize CSBEP}&\footnotesize(1,0.5')&\footnotesize96/59&\footnotesize40.45/38.91&\footnotesize2.0/3.6&\footnotesize1.63&\footnotesize1.22\\
--&\scriptsize \textbf{-South}&--&\footnotesize(1,1')&\footnotesize--/34&\footnotesize--/37.08&\footnotesize2.0/3.8&\footnotesize2.82&\footnotesize1.18\\
--&--&--&\footnotesize(0.3,0.5')&\footnotesize--/56&\footnotesize--/133.04&\footnotesize2.0/3.2&\footnotesize1.71&\footnotesize0.81\\
--&--&--&\footnotesize(0.3,1')&\footnotesize--/37&\footnotesize--/131.87&\footnotesize2.0/3.8&\footnotesize2.59&\footnotesize0.83\\
\textbf{\scriptsize ChaI2}&\textbf{\scriptsize ChaI}&\textbf{\scriptsize TNT}&\footnotesize(1,0.5')&\footnotesize96/67&\footnotesize40.45/42.38&\footnotesize1.4/3.2&\footnotesize1.43&\footnotesize1.27\\
--&\scriptsize \textbf{-South}&--&\footnotesize(1,1')&\footnotesize--/44&\footnotesize--/40.81&\footnotesize1.4/3.4&\footnotesize2.18&\footnotesize1.12\\
--&--&--&\footnotesize(0.3,0.5')&\footnotesize--/69&\footnotesize--/132.67&\footnotesize1.6/3.5&\footnotesize1.39&\footnotesize0.85\\
--&--&--&\footnotesize(0.3,1')&\footnotesize--/47&\footnotesize--/131.41&\footnotesize1.6/3.7&\footnotesize2.04&\footnotesize0.79\\
\textbf{\scriptsize ChaI3}&\textbf{\scriptsize ChaI}&\textbf{\scriptsize CSBET}&\footnotesize(1,0.5')&\footnotesize43/30&\footnotesize21.69/21.11&\footnotesize1.0/3.6&\footnotesize1.43&\footnotesize1.39\\
--&\scriptsize \textbf{-North}&--&\footnotesize(1,1')&\footnotesize--/18&\footnotesize--/20.22&\footnotesize1.0/3.5&\footnotesize2.39&\footnotesize1.16\\
--&--&--&\footnotesize(0.3,0.5')&\footnotesize--/27&\footnotesize--/70.42&\footnotesize1.0/3.0&\footnotesize1.59&\footnotesize0.89\\
--&--&--&\footnotesize(0.3,1')&\footnotesize--/21&\footnotesize--/70.21&\footnotesize1.0/4.0&\footnotesize2.05&\footnotesize0.79\\
\textbf{\scriptsize ChaI3}&\textbf{\scriptsize ChaI}&\textbf{\scriptsize CSBEP}&\footnotesize(1,0.5')&\footnotesize43/25&\footnotesize21.69/20.99&\footnotesize2.0/3.9&\footnotesize1.72&\footnotesize1.43\\
--&\scriptsize \textbf{-North}&--&\footnotesize(1,1')&\footnotesize--/13&\footnotesize--/20.00&\footnotesize2.0/4.6&\footnotesize3.31&\footnotesize1.26\\
--&--&--&\footnotesize(0.3,0.5')&\footnotesize--/25&\footnotesize--/71.45&\footnotesize2.0/3.7&\footnotesize1.72&\footnotesize1.03\\
--&--&--&\footnotesize(0.3,1')&\footnotesize--/16&\footnotesize--/70.83&\footnotesize2.0/3.9&\footnotesize2.69&\footnotesize0.94\\
\textbf{\scriptsize ChaI3}&\textbf{\scriptsize ChaI}&\textbf{\scriptsize TNT}&\footnotesize(1,0.5')&\footnotesize43/33&\footnotesize21.69/21.90&\footnotesize1.4/3.2&\footnotesize1.30&\footnotesize1.39\\
--&\scriptsize \textbf{-North}&--&\footnotesize(1,1')&\footnotesize--/19&\footnotesize--/20.96&\footnotesize1.4/3.3&\footnotesize2.26&\footnotesize1.14\\
--&--&--&\footnotesize(0.3,0.5')&\footnotesize--/34&\footnotesize--/69.99&\footnotesize1.6/3.3&\footnotesize1.26&\footnotesize1.00\\
--&--&--&\footnotesize(0.3,1')&\footnotesize--/19&\footnotesize--/69.19&\footnotesize1.6/3.7&\footnotesize2.26&\footnotesize0.96\\
\textbf{\scriptsize Lupus3}&\textbf{\scriptsize Lupus3}&\textbf{\scriptsize CSBET}&\footnotesize(1,0.5')&\footnotesize36/22&\footnotesize18.18/17.55&\footnotesize1.0/3.3&\footnotesize1.64&\footnotesize1.79\\
--&\scriptsize \textbf{-Main}&--&\footnotesize(1,1')&\footnotesize--/14&\footnotesize--/16.37&\footnotesize1.0/3.5&\footnotesize2.57&\footnotesize1.68\\
--&--&--&\footnotesize(0.3,0.5')&\footnotesize--/22&\footnotesize--/59.28&\footnotesize1.0/2.7&\footnotesize1.64&\footnotesize1.03\\
--&--&--&\footnotesize(0.3,1')&\footnotesize--/18&\footnotesize--/58.62&\footnotesize1.0/3.7&\footnotesize2.00&\footnotesize0.99\\
\textbf{\scriptsize Lupus3}&\textbf{\scriptsize Lupus3}&\textbf{\scriptsize CSBEP}&\footnotesize(1,0.5')&\footnotesize36/20&\footnotesize18.18/17.44&\footnotesize2.0/4.1&\footnotesize1.80&\footnotesize1.81\\
--&\scriptsize \textbf{-Main}&--&\footnotesize(1,1')&\footnotesize--/12&\footnotesize--/15.91&\footnotesize2.0/3.6&\footnotesize3.00&\footnotesize1.53\\
--&--&--&\footnotesize(0.3,0.5')&\footnotesize--/22&\footnotesize--/59.61&\footnotesize2.0/3.9&\footnotesize1.64&\footnotesize1.23\\
--&--&--&\footnotesize(0.3,1')&\footnotesize--/16&\footnotesize--/58.75&\footnotesize2.0/4.1&\footnotesize2.25&\footnotesize1.17\\
\textbf{\scriptsize Lupus3}&\textbf{\scriptsize Lupus3}&\textbf{\scriptsize TNT}&\footnotesize(1,0.5')&\footnotesize36/25&\footnotesize18.18/18.44&\footnotesize1.4/3.1&\footnotesize1.44&\footnotesize1.73\\
--&\scriptsize \textbf{-Main}&--&\footnotesize(1,1')&\footnotesize--/15&\footnotesize--/17.15&\footnotesize1.4/3.7&\footnotesize2.40&\footnotesize1.67\\
--&--&--&\footnotesize(0.3,0.5')&\footnotesize--/26&\footnotesize--/59.07&\footnotesize1.7/4.1&\footnotesize1.38&\footnotesize1.29\\
--&--&--&\footnotesize(0.3,1')&\footnotesize--/18&\footnotesize--/58.16&\footnotesize1.7/3.9&\footnotesize2.00&\footnotesize1.16\\
\textbf{\scriptsize IC3481}&\textbf{\scriptsize IC348}&\textbf{\scriptsize CSBET}&\footnotesize(1,0.5')&\footnotesize186/41&\footnotesize111.86/110.59&\footnotesize1.0/5.2&\footnotesize4.54&\footnotesize0.68\\
--&\scriptsize \textbf{-Main}&--&\footnotesize(1,1')&\footnotesize--/7&\footnotesize--/110.02&\footnotesize1.0/4.9&\footnotesize26.57&\footnotesize0.57\\
--&--&--&\footnotesize(0.3,0.5')&\footnotesize--/35&\footnotesize--/370.89&\footnotesize1.0/5.0&\footnotesize5.31&\footnotesize0.48\\
--&--&--&\footnotesize(0.3,1')&\footnotesize--/6&\footnotesize--/370.51&\footnotesize1.0/4.2&\footnotesize31.00&\footnotesize0.48\\
\textbf{\scriptsize IC3481}&\textbf{\scriptsize IC348}&\textbf{\scriptsize CSBEP}&\footnotesize(1,0.5')&\footnotesize186/32&\footnotesize111.86/110.63&\footnotesize2.0/5.2&\footnotesize5.81&\footnotesize0.75\\
--&\scriptsize \textbf{-Main}&--&\footnotesize(1,1')&\footnotesize--/6&\footnotesize--/110.11&\footnotesize2.0/5.7&\footnotesize31.00&\footnotesize0.53\\
--&--&--&\footnotesize(0.3,0.5')&\footnotesize--/21&\footnotesize--/371.93&\footnotesize2.0/7.0&\footnotesize8.86&\footnotesize0.61\\
--&--&--&\footnotesize(0.3,1')&\footnotesize--/5&\footnotesize--/371.36&\footnotesize2.0/4.7&\footnotesize37.20&\footnotesize0.47\\
\textbf{\scriptsize IC3481}&\textbf{\scriptsize IC348}&\textbf{\scriptsize TNT}&\footnotesize(1,0.5')&\footnotesize186/44&\footnotesize111.86/113.93&\footnotesize1.4/5.3&\footnotesize4.23&\footnotesize0.73\\
--&\scriptsize \textbf{-Main}&--&\footnotesize(1,1')&\footnotesize--/8&\footnotesize--/113.51&\footnotesize1.4/4.8&\footnotesize23.25&\footnotesize0.55\\
--&--&--&\footnotesize(0.3,0.5')&\footnotesize--/37&\footnotesize--/365.94&\footnotesize1.7/5.3&\footnotesize5.03&\footnotesize0.61\\
--&--&--&\footnotesize(0.3,1')&\footnotesize--/7&\footnotesize--/365.30&\footnotesize1.7/4.2&\footnotesize26.57&\footnotesize0.64\\
\textbf{\scriptsize IC3482}&\textbf{\scriptsize IC348}&\textbf{\scriptsize CSBET}&\footnotesize(1,0.5')&\footnotesize11/5&\footnotesize3.13/2.89&\footnotesize1.0/5.2&\footnotesize2.20&\footnotesize0.97\\
--&\scriptsize \textbf{-North}&--&\footnotesize(1,1')&\footnotesize--/1&\footnotesize--/2.62&\footnotesize1.0/4.9&\footnotesize11.00&\footnotesize--\\
--&--&--&\footnotesize(0.3,0.5')&\footnotesize--/4&\footnotesize--/10.22&\footnotesize1.0/5.0&\footnotesize2.75&\footnotesize0.80\\
--&--&--&\footnotesize(0.3,1')&\footnotesize--/2&\footnotesize--/10.00&\footnotesize1.0/4.2&\footnotesize5.50&\footnotesize0.79\\
\textbf{\scriptsize IC3482}&\textbf{\scriptsize IC348}&\textbf{\scriptsize CSBEP}&\footnotesize(1,0.5')&\footnotesize11/3&\footnotesize3.13/2.88&\footnotesize2.0/5.2&\footnotesize3.67&\footnotesize1.19\\
--&\scriptsize \textbf{-North}&--&\footnotesize(1,1')&\footnotesize--/1&\footnotesize--/2.60&\footnotesize2.0/5.7&\footnotesize11.00&\footnotesize--\\
--&--&--&\footnotesize(0.3,0.5')&\footnotesize--/3&\footnotesize--/10.19&\footnotesize2.0/7.0&\footnotesize3.67&\footnotesize0.90\\
--&--&--&\footnotesize(0.3,1')&\footnotesize--/2&\footnotesize--/9.97&\footnotesize2.0/4.7&\footnotesize5.50&\footnotesize0.76\\
\textbf{\scriptsize IC3482}&\textbf{\scriptsize IC348}&\textbf{\scriptsize TNT}&\footnotesize(1,0.5')&\footnotesize11/4&\footnotesize3.13/3.34&\footnotesize1.4/5.3&\footnotesize2.75&\footnotesize1.02\\
--&\scriptsize \textbf{-North}&--&\footnotesize(1,1')&\footnotesize--/1&\footnotesize--/3.08&\footnotesize1.4/4.8&\footnotesize11.00&\footnotesize--\\
--&--&--&\footnotesize(0.3,0.5')&\footnotesize--/5&\footnotesize--/9.98&\footnotesize1.5/5.3&\footnotesize2.20&\footnotesize0.76\\
--&--&--&\footnotesize(0.3,1')&\footnotesize--/2&\footnotesize--/9.77&\footnotesize1.5/4.2&\footnotesize5.50&\footnotesize0.79\\
\footnotetext[1]{\footnotesize (initial star formation efficiency, resolution)}
\footnotetext[2]{\footnotesize Total number of stars/ total number of derived cores and total mass of stars/ total mass of derived cores within each group \citep[stars data from][]{Kirk2010}}
\footnotetext[3]{\footnotesize Slope of M$\propto$R$^\beta$ for the input model/Slope of M$\propto$R$^\beta$ for the derived cores..}
\footnotetext[4]{ \footnotesize Crowding ratio f = mean separation/mean diameter}
\footnotetext[5]{\footnotesize Fragmentation ratio F = mean number of stars per observed core}
\end{longtable}

\begin{longtable}{ll|l}
\caption[Group divisions]{Blended groups} \label{tab:repartition} 
\\
\hline\hline\\[-3ex]
\multicolumn{1}{l}{\textbf{Model}}&
\multicolumn{1}{c}{\textbf{SFE 1}}&
\multicolumn{1}{c}{\textbf{SFE 0.3}}\\
\\[0.5ex]
\hline
\\[-1.8ex]
\endhead
\\[-1.8ex]\hline\hline
\endlastfoot
\textbf{CSBET}&IC 348-Main, IC 348-North&\small L1517, ChaI-South, ChaI-North, Lupus 3-Main,\\&&\small IC 348-Main, IC 348-North\\
\textbf{CSBEP}&IC 348-Main, IC 348-North&\small ChaI-South, ChaI-North, IC 348-Main, IC 348-North\\
\textbf{TNT}&IC 348-Main, IC 348-North&\small ChaI-South, ChaI-North, IC 348-Main, IC 348-North\\
\end{longtable}

\end{document}